\documentclass[a4paper,reprint,aps,prd,groupedaddress,superscriptaddress,nofootinbib]{revtex4-2}

\usepackage[T1]{fontenc}
\usepackage{lmodern}


\usepackage{amsmath,amssymb,mathtools}
\allowdisplaybreaks

\usepackage{siunitx}
\usepackage{float}

\usepackage{graphicx}
\usepackage{dcolumn}
\usepackage{bm}
\usepackage{physics}
\usepackage{booktabs}
\usepackage{tabularx}
\usepackage{multirow}
\usepackage{rotating}
\usepackage{adjustbox}
\usepackage[caption = false]{subfig}

\usepackage[normalem]{ulem}
\usepackage{comment}
\usepackage{csquotes}
\usepackage{orcidlink}

\PassOptionsToPackage{table}{xcolor}
\usepackage{xcolor}

\usepackage{microtype} 
\usepackage{hyperref}
\hypersetup{
  colorlinks=true,
  linkcolor=blue,
  citecolor=blue,
  urlcolor=blue
}
\usepackage[capitalise]{cleveref}

\begin{document}

\title{Negative neutrino mass or negative dark energy?}

\author{Cihad K{\i}br{\i}s\,\orcidlink{0000-0002-4129-2199}}
\email{kibrisc@itu.edu.tr}
\affiliation{Department of Physics, Istanbul Technical University, Maslak 34469 Istanbul, T\"{u}rkiye}

\author{Willem Elbers\,\orcidlink{0000-0002-2207-6108}}
\email{willem.h.elbers@durham.ac.uk}
\affiliation{Institute for Computational Cosmology, Department of Physics, Durham University, South Road, Durham, DH1 3LE, UK}

\author{\"{O}zg\"{u}r Akarsu\,\orcidlink{0000-0001-6917-6176}}
\email{akarsuo@itu.edu.tr}
\affiliation{Department of Physics, Istanbul Technical University, Maslak 34469 Istanbul, T\"{u}rkiye}

\author{Eleonora Di Valentino\,\orcidlink{0000-0001-8408-6961}}
\email{e.divalentino@sheffield.ac.uk}
\affiliation{School of Mathematical and Physical Sciences, University of Sheffield, Hounsfield Road, Sheffield S3 7RH, United Kingdom}


\begin{abstract}
Recent cosmological analyses based on DESI and CMB data have revealed a tension between the inferred sum of neutrino masses and the minimum value allowed by neutrino oscillation experiments, when assuming an underlying $\Lambda$CDM model of cosmology. In this work, we perform a systematic exploration of alternative dark energy models, including models that can supply a negative dark energy density capable of reproducing the cosmological effects of negative effective neutrino masses. We argue that dark energy models can alleviate the tension by modifying the cosmic expansion rate over a specific redshift range relevant for CMB lensing, while matching BAO distance measurements from DESI at lower redshifts. Among the models considered, we find that a sign-switching cosmological constant model, $\Lambda_\mathrm{s}$CDM, is uniquely capable of recovering positive neutrino masses by modifying the expansion history in this way. For the combination of DESI DR2 BAO, CMB, and DES-Dovekie supernova data, the constraint on the effective neutrino mass shifts from $\sum m_{\nu,\mathrm{eff}}=-0.075^{+0.039}_{-0.053}\,\mathrm{eV}$ (68\%) for $\Lambda$CDM to $\sum m_{\nu,\mathrm{eff}}=0.055\pm0.050\,\mathrm{eV}$ (68\%) for $\Lambda_\mathrm{s}$CDM, with a 95\% lower bound on the dark energy transition redshift, $z_\dagger>2.4$. Although $\Lambda_{\rm s}$CDM does not have the strongest overall statistical support among the models considered, when the $\sum m_{\nu,\mathrm{eff}}$ parameter is allowed to vary, our findings point toward a specific sign- and redshift-structured contribution to the late-time expansion history as a viable way to alleviate the neutrino mass tension.
\end{abstract}

\maketitle

\section{Introduction}

The existence of massive neutrinos provides concrete evidence for physics beyond the Standard Model. Over the past two decades, a sustained experimental effort has led to percent-level determinations of the mixing angles and mass splittings that characterize the standard three-neutrino framework~\cite{Capozzi:2017ipn,Capozzi:2018ubv,DeSalas:2018rby,Gonzalez-Garcia:2021dve,Capozzi:2021fjo,Esteban:2024eli,Capozzi:2025wyn}. Yet, while oscillation experiments impose a robust lower limit on the sum of the three known neutrino masses, $\sum m_\nu \geq \SI{0.059}{\eV}$~\cite{Esteban:2024eli}, their ordering and absolute mass scale remain uncertain.

Massive neutrinos play a nontrivial role in cosmology, affecting both the expansion history of the Universe and the growth of cosmic structure. These effects provide powerful and complementary means of probing neutrino masses independently of terrestrial experiments (see~\cite{Lesgourgues:2006nd,Wong:2011ip,Abazajian:2016hbv,Lattanzi:2017ubx,DiValentino:2024xsv} for reviews). In practice, the strongest constraints typically derive from measurements of the cosmic microwave background (CMB), combined with probes of the large-scale structure and late-time expansion history, particularly through baryon acoustic oscillations (BAO).

The latest cosmological observations have now reached a level of precision at which the inferred bounds on the sum of neutrino masses approach, and in some cases even surpass, the lower limits implied by oscillation experiments. In particular, BAO measurements from the Dark Energy Spectroscopic Instrument (DESI)~\cite{DESI:2025zpo,DESI:2025zgx}, when combined with the latest CMB data~\cite{Planck:2018vyg,ACT:2025fju,SPT-3G:2025bzu,SPT-3G:2025zuh}, provide the tightest constraints to date on $\sum m_\nu$. Conventional analyses of these data, adopting the standard $\Lambda$CDM model of cosmology, have produced 95\% upper limits in the range of $\sum m_\nu < 0.04-\SI{0.06}{\eV}$~\cite{Elbers:2025vlz,SPT-3G:2025bzu,Jiang:2024viw}, falling below the oscillation lower limit $\sum m_\nu \geq \SI{0.059}{\eV}$. This discrepancy, which we call \textit{the cosmological neutrino mass tension}, points to a failure of the underlying assumptions. If the tension is not a statistical fluctuation or a result of unidentified systematic errors, then it may signal new physics in the neutrino sector or a departure from the assumed $\Lambda$CDM cosmology.

The problem becomes more evident once the physical prior $\sum m_\nu \geq 0$ is relaxed and an effective neutrino-mass parameter is allowed to explore negative values. In this case, the likelihood distribution peaks in the unphysical region $\sum m_{\nu,\mathrm{eff}} < 0$~\cite{Craig:2024tky,Green:2024xbb,Elbers:2024sha,Elbers:2025vlz,Graham:2025dqn,Pulido-Hernandez:2026hcs,Yang:2026yaq,Loverde:2024nfi}, and the significance of the tension reaches the $3\sigma$ level. Intriguingly, a full-shape analysis of DESI galaxy clustering with limited prior information from the CMB also peaks in the range $\sum m_{\nu,\mathrm{eff}}<0$, albeit with increased uncertainty, and the same holds for analyses of CMB and CMB-lensing data alone~\cite{Elbers:2025vlz}. These findings suggest that the preferred cosmological solution may lie beyond the physical boundary imposed by the prior on the neutrino masses. However, the extension to $\sum m_{\nu,\mathrm{eff}}<0$ should only be considered as a diagnostic parameter, rather than as a proposal for negative neutrino rest masses.

The question we address is whether this diagnostic is pointing to a breakdown of the assumed cosmological model. Within $\Lambda$CDM, the sum of neutrino masses acts as a compensating parameter that can partially absorb mismatches among different cosmological observables, including BAO distance measurements and the growth and lensing information encoded in the CMB. Allowing this parameter to become negative effectively supplies a negative contribution to the effective cosmic energy budget governing the late-time Friedmann equation, thereby modifying both the expansion history and the growth of structure. From this perspective, the preference for $\sum m_{\nu,\mathrm{eff}}<0$ may indicate that the standard $\Lambda$CDM+$\sum m_\nu$ framework lacks relevant late-time degrees of freedom.

Recent analyses suggest a specific picture. The standard $\Lambda$CDM model struggles to simultaneously accommodate the shorter low-redshift BAO distance scales measured by DESI and the CMB lensing amplitude preferred by current data, while maintaining the precise geometric constraint from the angular acoustic scale in the CMB. At fixed CMB geometry, the shorter BAO distances pull the fit toward lower values of the fractional matter density, $\Omega_{\rm m}$, than those inferred from the CMB alone, but this in turn reduces the predicted CMB lensing amplitude. Adopting $\sum m_{\nu,\mathrm{eff}}<0$ can absorb this mismatch by reversing the usual cosmological effects of massive neutrinos: it lowers the effective late-time matter budget while enhancing the lensing signal. To replace $\sum m_{\nu,\mathrm{eff}}<0$ with a more compelling physical resolution, a model should either boost gravitational lensing directly or modify the expansion history~\cite{Graham:2025dqn}. In this study, we focus on models that modify the expansion history. We argue that these models should suppress $H(z)$ over the intermediate-redshift range most relevant for CMB lensing, while restoring a larger expansion rate at lower redshift so as to remain compatible with DESI BAO distances and CMB geometry.

Modifications in the dark energy (DE) sector, for example through dynamical dark energy (DDE) models, provide one possible source of such behavior. In low-dimensional equation-of-state (EoS)-based dynamical dark energy parametrizations, such as the Chevallier--Polarski--Linder (CPL) model~\cite{Chevallier:2000qy,Linder:2002et}, the neutrino-mass posterior can be broadened or shifted by altering the late-time expansion history~\cite{Elbers:2024sha,Elbers:2025vlz,Yang:2026yaq}. This is achieved by suppressing $H(z)$ during a high-redshift period with a phantom-like equation of state, $w_\mathrm{DE}(z)<-1$, before crossing the phantom-divide line (PDL) to a non-phantom period, $w_\mathrm{DE}(z)>-1$, at $z\lesssim0.5$. However, recovering a large positive value of $\sum m_\nu$ implies that $w_\mathrm{DE}(z)$ should evolve rather quickly, as illustrated, for example, in Fig.~3 of Ref.~\cite{Elbers:2024sha}. Additional constraints on the expansion history from Type Ia supernovae (SNIa) restrict rapidly evolving scenarios and therefore shift the posterior distribution of $\sum m_{\nu,\mathrm{eff}}$ back into the negative range.

We briefly comment on the terminology used above. For minimally coupled EoS-based parametrizations such as CPL, the DE density, $\rho_{\rm DE}$, remains sign-preserving once its present-day value is fixed to be positive. In this restricted positive-density setting, the null-energy-condition boundary (NECB), $\rho_{\rm DE}+p_{\rm DE}=0$, coincides with the usual phantom-divide line (PDL), $w_{\rm DE}=-1$~\cite{Akarsu:2026anp,Gokcen:2026pkq}. It is important, however, not to extrapolate this terminology too far. For density-level models in which $\rho_{\rm DE}$ can cross zero, the ratio $w_{\rm DE}=p_{\rm DE}/\rho_{\rm DE}$ may become ill-defined, and the physical NEC condition is controlled by $\rho_{\rm DE}+p_{\rm DE}=\rho_{\rm DE}(1+w_{\rm DE})$, not by whether $w_{\rm DE}$ lies above or below $-1$ alone~\cite{Akarsu:2026anp,Gokcen:2026pkq}. These points are addressed further in~\cref{model}.

The partial success of EoS-based models, such as CPL, points to a sharper interpretation of the anomaly. A viable late-time replacement for the phenomenological role played by $\sum m_{\nu,\mathrm{eff}}<0$ must generate a specific redshift-dependent deformation of the expansion history: suppressing $H(z)$ over the intermediate-redshift range where additional growth and CMB lensing are needed, while restoring a larger expansion rate at lower redshifts that is compatible with the DESI BAO distances and the CMB acoustic-scale constraint. Motivated by these considerations, in this work we investigate whether the preference for negative effective neutrino masses can be reinterpreted as a proxy for a missing negative effective contribution to the late-time expansion history. We analyze density-level dark energy models that can access a negative branch, including \textit{the sign-switching cosmological constant scenario}, $\Lambda_{\rm s}$CDM~\cite{Akarsu:2019hmw,Akarsu:2021fol,Akarsu:2022typ,Akarsu:2023mfb}, and the DMS20 realization of \textit{omnipotent dark energy} (ODE)~\cite{DiValentino:2020naf,Adil:2023exv,Specogna:2025guo}. We compare these with EoS-based benchmarks whose $\rho_{\rm DE}$ is sign-preserving, as well as with the geometric extension with curvature, $\Lambda$CDM+$\Omega_{\rm k}$, in order to distinguish generic broadening of parameter uncertainties from an explicit physical replacement for the role played by $\sum m_{\nu,\mathrm{eff}}<0$ in $\Lambda$CDM.

The sign-switching cosmological constant scenario is of particular interest because it provides a minimal and physically transparent realization of the required sign structure. In $\Lambda_{\rm s}$CDM, the effective vacuum-energy contribution is anti-de Sitter (AdS)-like (negative) for $z>z_\dagger$ and de Sitter (dS)-like (positive) for $z<z_\dagger$. Although the negative branch extends formally to earlier times, its main observational leverage is concentrated around the intermediate-redshift regime near the transition, while the model becomes nearly indistinguishable from $\Lambda$CDM at sufficiently high redshift. This makes $\Lambda_{\rm s}$CDM especially well suited as a benchmark for testing whether the neutrino-mass tension is pointing toward the specific kind of late-time deformation described above.

By confronting these models with current cosmological observations, including DESI BAO, CMB temperature and polarization data, CMB lensing reconstruction, and Type Ia supernova distances, we assess whether allowing a negative effective contribution to the late-time expansion history can restore physically consistent neutrino-mass constraints and, more broadly, clarify what kind of late-time freedom is being selected by the data. In this way, negative effective neutrino masses are recast as a concrete clue to missing structure in the late-time cosmological model.

The remainder of the paper is organized as follows. In~\cref{model}, we introduce the late-time extensions considered in the analysis. In~\cref{effect}, we review the background and perturbation-level effects of massive neutrinos. In~\cref{anatomy}, we describe the physical origin of the neutrino mass tension and its relation to DESI BAO and CMB lensing. The data sets and methodology are presented in~\cref{methods}. We report the parameter constraints in~\cref{results}, discuss their implications in~\cref{discussion}, and summarize our conclusions in~\cref{sec:conc}.

\section{Model framework} \label{model}

We distinguish two main classes of models, depending on the manner in which the dark energy is defined. The first class consists of models that specify directly the effective DE density, $\rho_{\rm DE}(a)$, allowing in particular for negative energy densities. This class includes the sign-switching cosmological constant scenario, $\Lambda_{\rm s}$CDM~\cite{Akarsu:2019hmw,Akarsu:2021fol,Akarsu:2022typ,Akarsu:2023mfb}, in which the effective vacuum-energy density is AdS-like (negative) for $z>z_\dagger$ and dS-like (positive) for $z<z_\dagger$, and the DMS20~\cite{DiValentino:2020naf} realization of omnipotent dark energy~\cite{Adil:2023exv,Specogna:2025guo}, whose non-monotonic density evolution can access a negative branch for suitable parameter choices.\footnote{This density-level motivation is also consistent with DESI-era model-agnostic, weakly parametric, and non-parametric reconstructions of the late-time expansion history and effective DE density. Using DESI BAO together with other datasets, including different Type~Ia supernova compilations, such analyses find indications that, with increasing redshift, the effective DE density can decrease rapidly toward very small values and, in some data combinations, approach or cross zero around the intermediate-redshift range $z\sim1.5$--2~\cite{DESI:2024aqx,Dinda:2024ktd,Sousa-Neto:2025gpj,Ormondroyd:2025exu,Berti:2025phi,DESI:2025fii,You:2025uon,Wang:2025vfb,Mukherjee:2025ytj,Gonzalez-Fuentes:2025lei,Akarsu:2026anp,Akarsu:2026pom}. Interpreted through the background continuity relation, this rapid decrease corresponds to NEC-violating, phantom-like behavior, $\rho_{\rm DE}(1+w_{\rm DE})<0$, and, while $\rho_{\rm DE}>0$, to $w_{\rm DE}<-1$; once $\rho_{\rm DE}$ approaches or crosses zero, however, the ratio $w_{\rm DE}=p_{\rm DE}/\rho_{\rm DE}$ becomes ill-conditioned, making the density-level description more appropriate. Related trends were also indicated by pre-DESI-era analyses, including studies based on SDSS BAO data~\cite{Holsclaw:2011wi,Seikel:2012uu,AlbertoVazquez:2012ofj,Zhao:2017cud,Wang:2018fng,Bonilla:2020wbn,Bernardo:2021cxi,Escamilla:2021uoj,Escamilla:2023shf,Escamilla:2024ahl,Sabogal:2024qxs}.} The second class consists of models that specify the equation of state (EoS) of DE. For these models, the sign of the energy density is fixed. This includes the $w$CDM model, with a constant equation of state, $w$, and the Chevallier--Polarski--Linder (CPL) parametrization of DE~\cite{Chevallier:2000qy,Linder:2002et}, together with various subcases introduced further below. Finally, we include $\Lambda$CDM+$\Omega_{\rm k}$ as a geometric benchmark, in which the background is modified through spatial curvature, $\Omega_{\rm k}$, rather than through the DE sector itself.

\textbf{\boldmath\textit{$\mathit{\Lambda}_{s}$CDM model}:} The sign-switching cosmological-constant scenario, $\Lambda_{\rm s}$CDM, was originally conjectured phenomenologically on the basis of findings from the \textit{graduated dark energy} (gDE) model~\cite{Akarsu:2019hmw}, which indicated that the presence of an effective AdS-like DE phase at earlier post-recombination times, followed by a sufficiently rapid transition to the present dS-like phase around $z\sim2$, could alleviate the $H_0$ tension and the discrepancy between CMB-inferred distances and BAO measurements from the Lyman-$\alpha$ forest of BOSS quasars reported at the time~\cite{Akarsu:2019hmw,Akarsu:2021fol,Akarsu:2022typ,Akarsu:2023mfb}. It posits that the vacuum-energy sector occupies an effective AdS-like branch, $\Lambda_{\rm AdS}<0$, for $z>z_\dagger$, and switches rapidly to the present dS-like branch, $\Lambda_{\rm dS}\simeq \abs{\Lambda_{\rm AdS}}>0$, for $z<z_\dagger$, with $z_\dagger\sim2$; the sign flip is approximately magnitude-preserving, in the sense of a \textit{mirror} AdS-to-dS-like transition. In this construction, the standard matter and radiation sectors, including baryons and CDM, as well as the inflationary initial conditions, are left unmodified; the pre-recombination calibration is preserved to a very good approximation because the vacuum-energy contribution is negligible at sufficiently high redshift. Such a transition can typically be described using sigmoid-like functions; for example, through the smooth approximation to the signum function, ${\rm sgn}\,x \simeq \tanh(kx)$ for a constant $k$, where $x$ is an appropriate transition variable, such as one defined in terms of the redshift $z$ or the scale factor $a=1/(1+z)$ in a Robertson--Walker spacetime~\cite{Akarsu:2022typ}. Accordingly, a representative realization of a sign-switching cosmological constant is
$\Lambda_{\rm s}(z)=\Lambda_{\rm dS}\tanh\!\left[\eta(z_\dagger-z)\right]$,
where $\eta>0$ controls the rapidity of the transition and $\Lambda_{\rm dS}$ sets the asymptotic dS magnitude.\footnote{At the level of the effective density history, one may associate this profile with
$\rho_{\Lambda_{\rm s}}(z)=\Lambda_{\rm s}(z)/(8\pi G)$. The corresponding effective inertial-mass density,
$\mathcal{I}_{\Lambda_{\rm s}}\equiv \rho_{\Lambda_{\rm s}}+p_{\Lambda_{\rm s}}$, follows from the background continuity relation as $\mathcal{I}_{\Lambda_{\rm s}}(z)
=
\frac{1+z}{3}\frac{{\rm d}\rho_{\Lambda_{\rm s}}}{{\rm d}z}$,
while the effective EoS parameter is $w_{\Lambda_{\rm s}}(z)
\equiv
p_{\Lambda_{\rm s}}/\rho_{\Lambda_{\rm s}}
=
-1+
\mathcal{I}_{\Lambda_{\rm s}}/\rho_{\Lambda_{\rm s}}(z)$. For the $\tanh$ profile, ${\rm d}\rho_{\Lambda_{\rm s}}/{\rm d}z<0$, so the inertial-mass density remains negative,
$\mathcal{I}_{\Lambda_{\rm s}}<0$, with its magnitude localized around the transition and approaching zero far from it, where the model tends to a cosmological-constant branch. Thus, the source violates the NEC, $\mathcal{I}_{\Lambda_{\rm s}}<0$, throughout the resolved transition and on both branches, while approaching the cosmological-constant limit, $\mathcal{I}_{\Lambda_{\rm s}}\to0$, far from the transition. However, because $\rho_{\Lambda_{\rm s}}$ changes sign, the ratio $w_{\Lambda_{\rm s}}$ satisfies $w_{\Lambda_{\rm s}}<-1$ on the positive-density branch, $z<z_\dagger$, and $w_{\Lambda_{\rm s}}>-1$ on the negative-density branch, $z>z_\dagger$, with a kinematical pole at $z=z_\dagger$, where $\rho_{\Lambda_{\rm s}}=0$~\cite{Ozulker:2022slu,Adil:2023exv,Paraskevas:2024ytz}. The apparent change in the sign of $1+w_{\Lambda_{\rm s}}$ is therefore not a conventional PDL crossing; it reflects the sign change of the density itself rather than a crossing of the NEC boundary, $\mathcal{I}_{\Lambda_{\rm s}}=0$~\cite{Akarsu:2025gwi,Akarsu:2026anp,Gokcen:2026pkq}. Far from the transition, ${\rm d}\rho_{\Lambda_{\rm s}}/{\rm d}z\to0$, and the model approaches a positive or negative cosmological constant on the corresponding branch.} For a sufficiently rapid transition, e.g.\ $\eta\gtrsim10$ with $z_\dagger\sim2$, one may safely take $\Lambda_{\rm dS}\simeq\Lambda_{\rm s0}\equiv\Lambda_{\rm s}(0)$. In the formal limit $\eta\rightarrow\infty$, one recovers what is commonly referred to in the literature as the \textit{abrupt} $\Lambda_{\rm s}$CDM model~\cite{Akarsu:2021fol,Akarsu:2022typ,Akarsu:2023mfb}, which provides a one-parameter extension of $\Lambda$CDM:
\begin{equation}
\Lambda_{\rm s}(z)\rightarrow \Lambda_{\rm s0}\,{\rm sgn}[z_\dagger-z]
\qquad \textnormal{for} \qquad \eta\rightarrow\infty,
\label{eq:ssdeff}
\end{equation}
where $\Lambda_{\rm s0}>0$ is the present-day value of $\Lambda_{\rm s}(z)$. This signum form should be understood as an idealized proxy for a rapid sign-switching transition, rather than as a claim that the underlying physical transition is fundamentally discontinuous. In this sense, the abrupt formulation is best viewed as a convenient limiting representation of the rapid-transition regime of the more general continuous framework.
 
From a mathematical point of view, $\Lambda_{\rm s}$CDM contains a positive, dS-like cosmological constant for $z<z_\dagger$, as in $\Lambda$CDM, while incorporating a negative, AdS-like cosmological constant for all redshifts prior to the transition, $z>z_\dagger$. Phenomenologically, however, when the transition lies in the observationally relevant range, $z_\dagger\sim2$, the observable impact of this modification is concentrated around the transition redshift, namely in the intermediate-redshift regime long after recombination. For $z<z_\dagger$, $\Lambda_{\rm s}$CDM closely follows the $\Lambda$CDM form of the expansion history, albeit with systematically different best-fitting background parameters and typically larger values of $H(z)$. Around the transition redshift, $z\sim z_\dagger$, the negative AdS-like branch produces a noticeable deformation of $H(z)$ relative to $\Lambda$CDM. At higher redshifts, the AdS-like vacuum-energy contribution rapidly becomes subdominant to matter and radiation, so the model becomes phenomenologically nearly indistinguishable from $\Lambda$CDM. Consequently, from an observational perspective, $\Lambda_{\rm s}$CDM functions as a post-recombination, late-time modification of the standard expansion history. It preserves the standard early-time calibration to a very good approximation, while the suppression of $H(z)$ above $z_\dagger$ is compensated by an enhanced low-redshift expansion rate below $z_\dagger$. This is precisely why cosmological fits in $\Lambda_{\rm s}$CDM naturally shift toward larger $H_0$ and lower $\Omega_{\rm m}$ relative to $\Lambda$CDM.

$\Lambda_{\rm s}$CDM has emerged as an economical and promising late-time extension of the standard $\Lambda$CDM model, motivated by the cosmological tensions that have persistently appeared over the past decade in the era of precision cosmology~\cite{Verde:2019ivm,DiValentino:2020zio,DiValentino:2021izs,Perivolaropoulos:2021jda,Schoneberg:2021qvd,Shah:2021onj,Abdalla:2022yfr,DiValentino:2022fjm,Kamionkowski:2022pkx,Giare:2023xoc,Hu:2023jqc,Verde:2023lmm,DiValentino:2024yew,CosmoVerseNetwork:2025alb,Ong:2025cwv,Akarsu:2024qiq}. A growing body of analyses indicates that, depending on the data combination considered, this framework can improve the fit relative to $\Lambda$CDM and alleviate several cosmological tensions and anomalies, including those associated with $H_0$, the closely related Type Ia supernova absolute magnitude $M_{\rm B}$, $S_8$, and the growth index $\gamma$~\cite{Akarsu:2019hmw,Akarsu:2021fol,Akarsu:2022typ,Akarsu:2023mfb,Paraskevas:2024ytz,Akarsu:2025ijk,Escamilla:2025imi,Akarsu:2025nns}, without simply transferring the mismatch to other sectors of the data. For instance, it leaves the age of the Universe compatible with estimates from the oldest globular clusters~\cite{Akarsu:2022typ}. Moreover, when physical neutrino-sector parameters are allowed to vary, the framework can remain compatible with standard expectations for $N_{\rm eff}$ and $\sum m_\nu$ while still alleviating cosmological tensions~\cite{Yadav:2024duq}. This last point is directly relevant to the present work, where the negative effective neutrino-mass direction is used not as a physical neutrino model, but as a diagnostic of missing late-time cosmological structure~\cite{Elbers:2024sha,Elbers:2025vlz}. The phenomenological success of $\Lambda_{\rm s}$CDM has also motivated significant theoretical progress toward dynamical realizations of the underlying mirror AdS-to-dS transition. Examples include the string-inspired $\Lambda_{\rm s}$CDM$^+$ constructions~\cite{Anchordoqui:2023woo,Anchordoqui:2024gfa,Anchordoqui:2024dqc,Soriano:2025gxd}, the type-II minimally modified-gravity realization $\Lambda_{\rm s}$VCDM~\cite{Akarsu:2024qsi,Akarsu:2024eoo,DeFelice:2020eju,DeFelice:2020cpt}, the teleparallel realization $f(T)$-$\Lambda_{\rm s}$CDM~\cite{Akarsu:2024nas,Souza:2024qwd}, the Ph-$\Lambda_{\rm s}$CDM model, which provides a smooth realization of the scenario within GR through a phantom scalar field evolving on a suitable potential~\cite{Akarsu:2025gwi,Akarsu:2025dmj,Adil:2026kfn}, and, in certain formulations of GR, a construction based on an overall sign change of the metric~\cite{Alexandre:2023nmh}. In the present analysis, we use the abrupt $\Lambda_{\rm s}$CDM limit introduced in~\cref{eq:ssdeff} as the minimal benchmark realization of a sign-switching vacuum-energy history.

\textbf{\textit{DMS20 model of omnipotent dark energy}:} Omnipotent dark energy refers to a class of DE models characterized by non-monotonic density histories capable of accessing a negative-density branch~\cite{Adil:2023exv}. In such models, the effective equation-of-state ratio may exhibit phantom-divide behavior and, when $\rho_{\rm DE}$ crosses zero, apparent singularities in $w_{\rm DE}=p_{\rm DE}/\rho_{\rm DE}$; these features should be understood as properties of the ratio $w_{\rm DE}$, rather than as singularities of the background expansion. The DMS20 parametrization, first introduced in Ref.~\cite{DiValentino:2020naf}, provides a concrete realization of this broader idea. Recent work~\cite{Adil:2023exv}, which updated and extended the data sets used in the original study, reaffirmed the effectiveness of DMS20 in mitigating observational discrepancies and clarified the important role played by its ability to access a negative-density branch, a feature that had previously received less attention than the phantom-crossing behavior built into the parametrization. The DMS20 model parametrizes $\rho_{\rm DE}$ so as to ensure an extremum at a scale factor $a_\mathrm{m}$, where $\eval{\dd{\rho_{\rm DE}}/\dd{a}}_{a=a_\mathrm{m}}=0$, with the DE density expressed as
\begin{equation}
 \rho_{\rm DE}(a)=\rho_{\rm DE0}\frac{1+\alpha (a-a_\mathrm{m})^2 +\beta (a-a_\mathrm{m})^3}{1+\alpha (1-a_\mathrm{m})^2 +\beta (1-a_\mathrm{m})^3},
 \label{eq:rhoExp}
\end{equation}
where $\alpha$ and $\beta$ are constants associated with the quadratic and cubic departures from a cosmological constant --- see Refs.~\cite{DiValentino:2020naf,Adil:2023exv,Specogna:2025guo} for details. 
The cubic term governed by $\beta$ allows the density history, within the parameter domain explored here, to pass from a negative-density branch in the past to a positive-density branch at later times. Using the same background relation discussed for the smooth $\Lambda_{\rm s}$CDM profile, the effective inertial-mass density is
$\mathcal{I}_{\rm DE}\equiv\rho_{\rm DE}+p_{\rm DE}=-(a/3)\,{\rm d}\rho_{\rm DE}/{\rm d}a$, and the corresponding EoS ratio is
$w_{\rm DE}=-1+\mathcal{I}_{\rm DE}/\rho_{\rm DE}$. Hence, when $\rho_{\rm DE}$ crosses zero, $w_{\rm DE}=p_{\rm DE}/\rho_{\rm DE}$ develops a kinematical pole~\cite{Ozulker:2022slu,Adil:2023exv}. For the negative-to-positive density transition realized in the parameter range used here, one has $w_{\rm DE}>-1$ on the negative-density side and $w_{\rm DE}<-1$ immediately after entering the positive-density branch, in the same sense as for the sign-changing density history of the smooth $\Lambda_{\rm s}$CDM proxy. This behavior is not, by itself, a conventional phantom-divide-line crossing; it is caused by the sign change of $\rho_{\rm DE}$~\cite{Akarsu:2025gwi,Akarsu:2026anp,Gokcen:2026pkq}.

The DMS20 model can also realize genuine PDL crossings at lower redshift, after the DE density has become positive. These occur at extrema of $\rho_{\rm DE}$ for which ${\rm d}\rho_{\rm DE}/{\rm d}a=0$ and $\rho_{\rm DE}\neq0$, so that $\mathcal{I}_{\rm DE}=0$ and $w_{\rm DE}=-1$. In the DMS20 parametrization, this includes the point $a=a_{\rm m}$, and, when it lies in the physical domain, the additional crossing $a_\mathrm{n}=a_\mathrm{m}-2\alpha/(3\beta)$~\cite{Adil:2023exv}. In the parameter range adopted here, the relevant lower-redshift PDL crossing occurs within the positive-density regime and gives the conventional transition between phantom-like and non-phantom-like EoS behavior. Thus, DMS20 combines two distinct ingredients: a smooth negative-to-positive density transition in the past, and subsequent conventional PDL-crossing behavior associated with non-monotonic positive-density evolution~\cite{Adil:2023exv,Specogna:2025guo}.

This makes DMS20 useful in the present context, where the relevant issue is not generic late-time freedom alone, but the ability to supply a negative contribution to the effective cosmic energy budget in a way that modifies the expansion history over the redshift range relevant to the anomaly, and provides a representative ODE benchmark against which the more minimal $\Lambda_{\rm s}$CDM construction can be compared.

\textbf{\textit{EoS-based models}:}
We also consider various commonly used EoS-based parametrizations of the dark energy sector. The constant-$w$ model ($w$CDM) is the simplest parametrization in this class, recovering $\Lambda$CDM in the $w=-1$ limit. The two-parameter CPL family of models, in which the equation of state is given by $w_{\rm DE}(a)=w_0+w_a(1-a)$, allows for a richer phenomenology with an evolving equation-of-state parameter~\cite{Chevallier:2000qy,Linder:2002et}. Within this class, we further single out two theoretically motivated restrictions: non-phantom dynamical dark energy (NPDDE), defined by the condition $w_{\rm DE}(a)\geq -1$ over the relevant domain, and Mirage DE, corresponding to the trajectory $w_a=-3.66(1+w_0)$~\cite{Linder:2007ka}. By construction, NPDDE excludes phantom crossing. Mirage dark energy, meanwhile, is motivated by the observation that certain evolving DE trajectories can mimic $\Lambda$CDM with $w\approx-1$ by preserving approximately the $\Lambda$CDM distance to the last-scattering surface.

For these low-dimensional EoS-based dark energy models, $\rho_{\rm DE}(a)$ is determined by the continuity equation, assuming that the DE component is separately conserved, as appropriate for minimally coupled DE: $\dot{\rho}_{\rm DE}+3H\left[1+w_{\rm DE}(a)\right]\rho_{\rm DE}=0$.
This gives
\begin{equation}
\rho_{\rm DE}(a)=\rho_{\rm DE0}
\exp\!\left[-3\int_1^a \frac{1+w_{\rm DE}(a')}{a'}\,{\rm d}a'\right].    
\end{equation}
For the CPL form, this reduces to
\begin{equation}
\rho_{\rm DE}(a)/\rho_{\rm DE0}
= a^{-3(1+w_0+w_a)}\exp[-3w_a(1-a)].    
\end{equation}
The multiplicative factor relating $\rho_{\rm DE}(a)$ to $\rho_{\rm DE0}$ is strictly positive for finite and non-singular real $w_{\rm DE}(a)$. Therefore, for the minimally coupled EoS-based models considered here, fixing $\rho_{\rm DE0}>0$ fixes the sign of $\rho_{\rm DE}(a)$ throughout the evolution. In this sense, it is the DE density, not the EoS parameter itself, that is sign-preserving; $w_{\rm DE}(a)$ may still evolve and, in generic CPL, may cross $w=-1$. These models can therefore redistribute the late-time distance budget, but they cannot realize a negative-density branch~\cite{Akarsu:2026pom}. In what follows, this is what we mean by EoS-based models with sign-preserving $\rho_{\rm DE}$.

\textbf{\textit{Spatial curvature}:} Finally, as a more conventional modification of the expansion history, we consider the minimal geometric extension of $\Lambda$CDM with constant spatial curvature, $\Lambda$CDM+$\Omega_{\rm k}$. Spatial curvature enters the dimensionless Friedmann equation, $E^2(z)\equiv H^2(z)/H_0^2$, as $\Omega_{\rm k}(1+z)^2$ and can therefore contribute with either sign; however, the sign of the curvature contribution itself is fixed throughout cosmic history, and it carries no independent transition scale. For $\Omega_{\rm k}<0$, corresponding to a spatially closed universe, curvature acts as a negative contribution to $E^2(z)$. Moreover, if one formally groups it with the positive cosmological constant, the combined effective contribution
$\Omega_\Lambda+\Omega_{\rm k}(1+z)^2$
can cross zero at
$z_{\rm cross}=\sqrt{\Omega_\Lambda/|\Omega_{\rm k}|}-1$,
provided this crossing occurs at positive redshift. In this limited sense, closed $\Lambda$CDM provides the simplest geometric analogue of a sign-changing effective contribution to the Friedmann equation~\cite{Acquaviva:2021jov}. This analogy is useful, but should not be overinterpreted: unlike $\Lambda_{\rm s}$CDM or DMS20, curvature does not describe a sign-switching DE density, and it also modifies the distance-redshift relations geometrically. It therefore serves here as a geometric control rather than as a genuine sign-switching DE history.

\section{The Effect of Neutrinos on Cosmological Dynamics} \label{effect}

Massive neutrinos affect cosmological dynamics at both the background and perturbation levels. Since the anomaly studied in this work is tied simultaneously to late-time distance measurements and to the CMB lensing amplitude, it is useful to briefly recall how neutrino mass enters each of these sectors. We follow the exposition in~\cite{Elbers:2025vlz} and refer to~\cite{Lesgourgues:2006nd,Wong:2011ip,Abazajian:2016hbv,Lattanzi:2017ubx,DiValentino:2024xsv} for comprehensive reviews. Throughout this section, we keep $c$ explicit in distance and lensing-projection formulae, while using the standard cosmological convention $c=\hbar=k_{\rm B}=1$ in thermal and energy-density expressions.

\subsection{Background Level}

Neutrinos behave as radiation at early times and as matter at late times. As the Universe expands, they transition from radiation-like to matter-like behavior around the epoch when the neutrino temperature becomes comparable to the rest mass of a given eigenstate, $T_\nu \sim m_i$. At the background level, their energy density is well approximated by
\begin{equation}
\label{Neutrinobehavior}
    \rho_\nu(a) =  \displaystyle \sum_{i = 1}^{N_\nu} \dfrac{T_{\nu}^4(a)}{\pi^2} \displaystyle \int_{0}^{\infty} \dfrac{x^2 \sqrt{x^2 + [m_i/ T_\nu(a)]^2}}{1 + e^x}{\rm d}x\, ,
\end{equation}
where $T_\nu(a) = T_{\nu,0}/a$, with $T_{\nu,0}$ being the present-day temperature of neutrinos. To a very good approximation, we can split~\cref{Neutrinobehavior} into two limiting regimes, $z \gg z_{\rm nr}$ and $z \ll z_{\rm nr}$, where $z_{\rm nr} \approx 1890\,(m_i / {\rm eV})$ is the redshift of the relativistic-to-nonrelativistic transition.
For illustration, assuming the minimal total mass under the normal mass ordering, $\sum m_\nu\simeq\SI{0.06}{\eV}$, the heaviest mass eigenstate with $m_i\simeq\SI{0.05}{\eV}$ would undergo this transition at $z_{\rm nr}\simeq 95$. In general, for the sub-eV masses relevant for our discussion, the transition would occur well after recombination. By contrast, a neutrino eigenstate with $m_i\simeq0.58\,{\rm eV}$ becomes nonrelativistic around recombination. In the relativistic and nonrelativistic limits, the neutrino energy density is given by
\begin{align}
&\text{at}\;z\gg z_{\rm nr}, \quad \rho_\nu(z) = N_{\rm eff} \dfrac{7 \pi^2}{120} T_{\nu,0}^4 \,(1+z)^4\, , \\
&\text{at}\;z\ll z_{\rm nr}, \quad \rho_\nu(z) = \dfrac{\sum m_\nu}{93.14\,{\rm eV}\,h^2}\dfrac{3 H_0^2}{8 \pi G}(1 + z)^3\, ,
\label{eq:rho_nu_limits}
\end{align}
where $N_{\rm eff}$ is the effective number of relativistic neutrino species and $\sum m_\nu$ is the sum of neutrino masses.

A useful quantity for understanding the background effects is the comoving angular diameter distance to recombination, given by
\begin{equation}
\label{angular_diameter_dist}
    D_{\rm M}(z_\star) = \int_0^{z_\star}\dfrac{c\,{\rm d}z}{H(z)}\, ,
\end{equation}
in the spatially flat case, where $z_\star$ is the redshift of recombination. For later use, we define the comoving radial distance,
$\chi(z)\equiv \int_0^z c\,dz'/H(z')$, so that $D_{\rm M}(z)=\chi(z)$ in the spatially flat case.

The acoustic oscillations in the CMB determine a characteristic angular scale, $\theta_\star = r_\star/D_{\rm M}(z_\star)$, where $r_\star$ is the sound horizon at recombination. The value of $\theta_\star$ is pinned down precisely by the spacing of the acoustic peaks in the CMB power spectra, corresponding to multipoles $\ell\simeq m\ell_A$ with
\begin{equation}
\label{multipole2}
\ell_A \simeq \pi \dfrac{D_{\rm M}(z_\star)}{r_\star}.
\end{equation}
For the sub-eV neutrino masses under consideration, neutrinos remain effectively relativistic before recombination, so the standard pre-recombination calibration depends only very weakly on the neutrino mass. Therefore, the precise measurement of $\theta_\star$ also tightly constrains $D_{\rm M}(z_\star)$, since $r_\star$ is determined by pre-recombination physics.

Once neutrinos become nonrelativistic, however, they contribute to the matter density, with $\Omega_\nu h^2 = \sum m_\nu/93.14\,{\rm eV}$, thereby increasing the late-time expansion rate. The faster expansion at $z<z_{\rm nr}$ reduces $D_{\rm M}(z_\star)$ and shifts the acoustic peaks toward larger scales and lower multipoles, as can be seen from~\cref{multipole2}. The precise alignment with the observed CMB can then typically be restored by reducing the late-time expansion rate, which in practice corresponds to a lower value of the Hubble constant, $H_0$. This is the well-known geometric degeneracy between $H_0$ and $\sum m_\nu$~\cite{Lesgourgues:2006nd,Lesgourgues2013}.

\subsection{Perturbation Level}

Neutrinos are produced with high thermal velocities and therefore free-stream in and out of matter overdensities. This velocity dispersion makes it difficult to confine them within CDM potential wells, rendering their clustering inefficient on scales smaller than their free-streaming length. Assuming that the neutrinos become nonrelativistic during the matter-dominated era, the characteristic wavenumber associated with the free-streaming scale at the transition is approximately
\begin{equation}
    k_{\rm nr} \simeq 0.018 \sqrt{\Omega_{\rm m}} \bigg( \dfrac{m_i}{1\, \rm eV}\bigg)^{1/2} \, h/\rm Mpc\, ,
\end{equation}
where $m_i$ corresponds to the mass of each neutrino mass eigenstate~\cite{Lesgourgues:2012uu}.
On large scales, $k\ll k_{\rm nr}$, neutrinos cluster more efficiently and contribute to the growth of structure, similarly to CDM and baryons, albeit with minor relativistic corrections.
On scales $k\gg k_{\rm nr}$, by contrast, free-streaming suppresses the growth of neutrino perturbations. However, neutrinos still contribute fully to the background expansion and, as a result, the growth of CDM and baryon perturbations is also suppressed. Overall, this scale-dependent suppression of clustering leaves a distinctive imprint in the matter power spectrum, $P_\delta(k)$, which, in the linear regime and on sufficiently small scales, is approximated by
\begin{equation}
    \dfrac{\Delta P_\delta(\sum m_\nu)}{P_\delta(\sum m_\nu = 0)} \simeq -8f_\nu\, ,
\end{equation}
where $f_\nu \equiv \Omega_\nu/\Omega_{\rm m}$ is the neutrino mass fraction~\cite{Hu:1997mj,Kiakotou:2007pz,Lesgourgues:2006nd}. On nonlinear scales, the maximum effect is even greater, $\Delta P_\delta/P_\delta\simeq -10f_\nu$~\cite{Brandbyge:2008rv,Viel:2010bn}. For realistic neutrino masses, say around $\sum m_\nu=0.10\,{\rm eV}$, the amplitude of the matter power spectrum is therefore suppressed by several percent.

Detecting the imprint of neutrino free streaming remains a major outstanding challenge for observational cosmology. One of the obstacles is that $P_\delta(k)$ is not observed directly. Galaxy redshift surveys probe the matter field through the number density and clustering of galaxies, but galaxies are biased tracers of the underlying large-scale structure. On large scales, the galaxy and matter power spectra are related by $P_g(k,z)=b^2P_\delta(k,z)$, where $b$ is the linear bias factor. Current constraints on the neutrino mass from free streaming therefore primarily derive from the shape of the galaxy power spectrum, rather than its amplitude~\cite{Brieden:2022lsd,Elbers:2025vlz}. Measurements of the galaxy bispectrum offer a promising way to break this degeneracy~\cite{Chudaykin:2019ock,Hahn:2019zob}.

Weak-lensing measurements, including cosmic shear and CMB lensing, provide a complementary probe of the matter distribution, as photon trajectories are distorted by gravitational potentials sourced by the total matter distribution along the line of sight. Since massive neutrinos inhibit the growth of matter perturbations on scales $k>k_{\rm nr}$, they reduce the corresponding lensing amplitude. Weak lensing of the CMB, which originates from a precisely known source plane at $z_\star\simeq1090$~\cite{Lewis:2006fu}, is one of the most powerful observables for probing neutrino mass. Measurements of CMB lensing, both through the smoothing of the CMB temperature and polarization spectra and especially through reconstruction of the lensing potential power spectrum from the four-point function, therefore play a key role in constraining the sum of neutrino masses.

\section{Anatomy of the Tension} \label{anatomy}

A growing body of work indicates that the neutrino mass tension is largely driven by two factors. The first is a discrepancy between the low-redshift BAO distance ratios measured by DESI and the $\Lambda$CDM expectation inferred from CMB data. More specifically, DESI favors smaller transverse and line-of-sight distance ratios, namely, $D_{\rm M}(z)/r_{\rm d}$ and $D_{\rm H}(z)/r_{\rm d}$, where $D_{\rm H}(z)\equiv c/H(z)$ and $r_{\rm d}$ is the sound horizon at the drag epoch. Independently of any early-time calibration, this corresponds to a preference for larger $H_0 r_{\rm d}$ compared to the CMB expectation. Within $\Lambda$CDM, such an increase in $H_0 r_{\rm d}$ is accompanied by a lower fractional matter density, $\Omega_{\rm m}$, than that inferred from the CMB (and supernova samples), leading to a discordance of $2$--$3\,\sigma$ in the preferred values of $\Omega_{\rm m}$. Therefore, when the effective neutrino-mass parameter $\sum m_{\nu,\mathrm{eff}}$ is allowed to vary in $\Lambda$CDM, combining CMB data with DESI BAO shifts the posterior along the geometric degeneracy direction, favoring lower $\Omega_{\rm m}$ and driving $\sum m_{\nu,\mathrm{eff}}$ to smaller values, with its preferred region extending into the negative domain. For example, along the CMB-only geometric degeneracy, matching the DESI-preferred value of $H_0 r_{\rm d}$ requires an effective neutrino mass as negative as $\sum m_{\nu,\rm eff}\simeq -0.11\,{\rm eV}$~\cite{Elbers:2025vlz}.

In addition to these background-level degeneracies, the neutrino-mass tension receives an important contribution from the preference of \emph{Planck} CMB data for excess gravitational lensing~\cite{Planck:2018vyg,Mccarthy:2017yqf,RoyChoudhury:2019hls,DiValentino:2019dzu,Allali:2024aiv,Craig:2024tky,Elbers:2024sha,Naredo-Tuero:2024sgf,Green:2024xbb}. This can emerge both as extra peak smoothing in the two-point temperature and polarization spectra, often captured by a phenomenological rescaling of the lensing potential $C_\ell^{\phi\phi}\!\to\!A_{\rm lens}C_\ell^{\phi\phi}$~\cite{Calabrese:2008rt}, with $A_{\rm lens}>1$, and, depending on the adopted likelihood, as an enhanced lensing potential power spectrum, $C_\ell^{\phi\phi}$, inferred directly from CMB four-point (trispectrum) measurements, sometimes quantified by a separate reconstruction amplitude, $A^{\phi\phi}_{\rm lens}$.

Typically, combining DESI BAO with CMB data strengthens any pre-existing preference for $A_{\rm lens}>1$ and, depending on the adopted likelihood combination, can even induce such a preference in the joint analysis by pulling the fit toward the lower-$\Omega_{\rm m}$ region favored by DESI. Given DESI's preference for lower $\Omega_{\rm m}$ values and the lensing anomaly, which this preference can further exacerbate, any model capable of enhancing the predicted CMB lensing amplitude while remaining compatible with the low-redshift BAO distance ratios preferred by DESI can be expected to shift the posterior distribution of $\sum m_{\nu,\rm eff}$ toward the positive domain.

One way to realize both requirements is to reduce the expansion rate $H(z)$ over the intermediate-redshift range $1.5 \lesssim z \lesssim 10$, where the CMB lensing kernel carries substantial weight, while increasing $H(z)$ again at lower redshift so as to accommodate the shorter DESI BAO distances relative to $\Lambda$CDM. Within the minimal phenomenological extension $\Lambda$CDM$+\sum m_{\nu,\rm eff}$, allowing $\sum m_{\nu,\rm eff}<0$ is an economical way to generate this pattern in $H(z)$ while preserving the early-time calibration. Since such values are not physically admissible, however, they should be interpreted as a signal that the fit is seeking additional late-time freedom capable of simultaneously reproducing the DESI BAO distances and the preferred level of CMB lensing. A more physical realization may instead be provided by DE density histories that supply a negative effective contribution to the expansion history over the relevant intermediate-redshift range, such as a sign-switching cosmological constant~\cite{Akarsu:2019hmw,Akarsu:2021fol,Akarsu:2022typ,Akarsu:2023mfb} or more general omnipotent DE scenarios~\cite{DiValentino:2020naf,Adil:2023exv,Specogna:2025guo}. When such a density history suppresses $H(z)$ over the lensing-sensitive intermediate-redshift interval, it reduces the effective Hubble friction, enhances structure growth, and thereby increases the corresponding CMB lensing signal.
This enhancement can be quantified through the lensing potential power spectrum, $C^{\phi\phi}_{\ell}$, which, in the Limber approximation, reads
\begin{equation}\label{eqn:cphiphi}
    C^{\phi\phi}_\ell \approx \dfrac{8\pi^2}{\ell^3}\displaystyle \int_0^{\chi_\star} \chi {\rm d}\chi \bigg(\dfrac{\chi_\star - \chi}{\chi_\star\chi} \bigg)^2 {\cal P}_\Psi\!\bigl(k,z(\chi)\bigr).
\end{equation}
Here, ${\cal P}_\Psi(k,z)$ is the dimensionless power spectrum of the Weyl potential, $\chi(z)$ is the comoving radial distance defined above, and $k\simeq(\ell+1/2)/\chi$ under the Limber approximation~\cite{LoVerde:2008re}. We can relate ${\cal P}_\Psi(k,z)$ to the usual matter power spectrum $P_\delta(k,z)$ via the Poisson equation as
\begin{equation}\label{eqn:phipsi}
    {\cal P}_\Psi(k, z) = \dfrac{9\Omega_{\rm m}^2H_0^4}{8\pi^2 c^4\,k}(1+z)^2 P_\delta(k,z).
\end{equation}
CMB lensing analyses commonly report the convergence power spectrum, $C^{\kappa\kappa}_\ell$, which, in terms of $C^{\phi\phi}_\ell$, can be written as $C_\ell^{\kappa\kappa} = \ell^2(\ell+1)^2C_\ell^{\phi\phi}/4$~\cite{Lewis:2006fu}, leading to
\begin{equation}
\label{convergence}
    \!\!C^{\kappa\kappa}_\ell \!\approx\! \dfrac{9 \Omega_{\rm m}^2H_0^4}{4c^4}\displaystyle\int_0^{\chi_\star}{\rm d}\chi (1+z)^2 \bigg (1-\dfrac{\chi}{\chi_\star}\bigg)^2 P_\delta [k;z(\chi)].
\end{equation}
$P_\delta(k,z)$ is the linear matter power spectrum, which schematically takes the form
\begin{equation}
\label{eqn:pdelta}
    P_\delta(k,z) = {\cal A}_s \dfrac{8\pi^2 c^4}{25\Omega_{\rm m}^2 H_0^4} \bigg( \dfrac{k}{k_{\rm p}} \bigg)^ {n_\mathrm{s} - 1} k \, D_+^2(k,z) T^2(k),
\end{equation}
where ${\cal A}_s$ is the amplitude of the primordial scalar perturbations, $n_\mathrm{s}$ is the spectral index, $k_{\rm p}$ is the pivot scale, often taken to be $0.05\, \rm Mpc^{-1}$, $D_+(k,z)$ is the (unnormalized) linear growth factor, and $T(k)$ is the transfer function that approaches unity for small $k$ values. Neglecting the scale-dependent growth introduced by the presence of massive neutrinos, $D_+(k,z)\approx D_+(z)$ can be obtained by solving the linear growth equation
\begin{equation}
    \label{lineargrowthequaion}
    D_+''+\bigg(\dfrac{3}{a}+\dfrac{E'}{E}\bigg)D_+'-\dfrac{3}{2}\bigg(\dfrac{\Omega_{\rm m}}{a^5 E^2}\bigg)D_+ = 0,
\end{equation}
where $E \equiv E(a) := H(a)/H_0$ is the normalized Hubble parameter and the prime~($\prime$) represents the derivative with respect to $a$, the scale factor of the Universe.
For fixed $\Omega_{\rm m}h^2$ and $\theta_\star$, DE models that reduce the expansion rate at intermediate redshifts tend to reduce the effective Hubble-friction term in~\cref{lineargrowthequaion}, which in turn enhances the growth of structure and hence the lensing amplitude.

Moreover, the modified background geometry can slightly increase lensing efficiency by reweighting the convergence integral~\cref{convergence} and, depending on scale, by shifting the projection through $k\simeq(\ell+1/2)/\chi$. Overall, the resulting enhancement of the lensing signal can help relieve the neutrino-mass tension to some extent.

Equations~\eqref{eqn:cphiphi}--\eqref{convergence} summarize the standard lensing-projection relations in the Limber approximation and are used here to motivate how changes in the background geometry feed into the CMB lensing signal. To build further intuition for the sign of the growth response, we also invoke the simplified linear relations in Eqs.~\eqref{eqn:pdelta} and~\eqref{lineargrowthequaion}. In the presence of massive neutrinos, however, the actual growth is scale dependent, so Eqs.~\eqref{eqn:pdelta} and~\eqref{lineargrowthequaion} should be read as heuristic guidance rather than as the numerical model used in the analysis. The representative spectra shown in~\cref{fig:fractional_lens} are computed numerically with the Boltzmann solver \texttt{CAMB}~\cite{Lewis:1999bs}, while the parameter inference reported in~\cref{results} uses a modified version of \texttt{CLASS}~\cite{Lesgourgues:2011re,Lesgourgues:2011rh} introduced in~\cref{methods}.

\paragraph*{What kind of late-time modification can replace a negative effective neutrino mass?}
Based on the discussion above, we summarize the requirements that a DE model should satisfy in order to alleviate the neutrino mass tension. At the background level, the model must satisfy two conditions. For fixed $\Omega_{\rm m} h^2$ and the accurately measured acoustic scale, $\theta_\star$, it should reduce $H(z)$ over the intermediate-redshift range $1.5 \lesssim z \lesssim 10$, where the CMB lensing kernel carries substantial weight, while increasing $H(z)$ again at lower redshifts so as to remain compatible with the shorter DESI BAO distances.

The perturbation-level requirement is equally important. From the above expressions for $C_\ell^{\kappa\kappa}$, $P_\delta(k,z)$, and the linear growth equation, lowering $H(z)$ over the relevant intermediate-redshift interval reduces the Hubble-friction term in the growth equation and thereby enhances $D_+(z)$ relative to $\Lambda$CDM. Through the combined effects of enhanced matter clustering and the modified background geometry entering the convergence integral, this tends to increase the predicted CMB lensing power. In this sense, a viable physical replacement for $\sum m_{\nu,\mathrm{eff}}<0$ must not only reproduce the required late-time expansion history, but also provide the additional growth needed to accommodate the observed lensing excess.

We now turn to several concrete realizations of this logic within the late-time extensions considered below.

\subsection{Neutrino Mass in \boldmath$\Lambda_{\rm s} \rm CDM$}

\begin{figure}[hb!]
    \centering
    \includegraphics[width=0.48\textwidth]{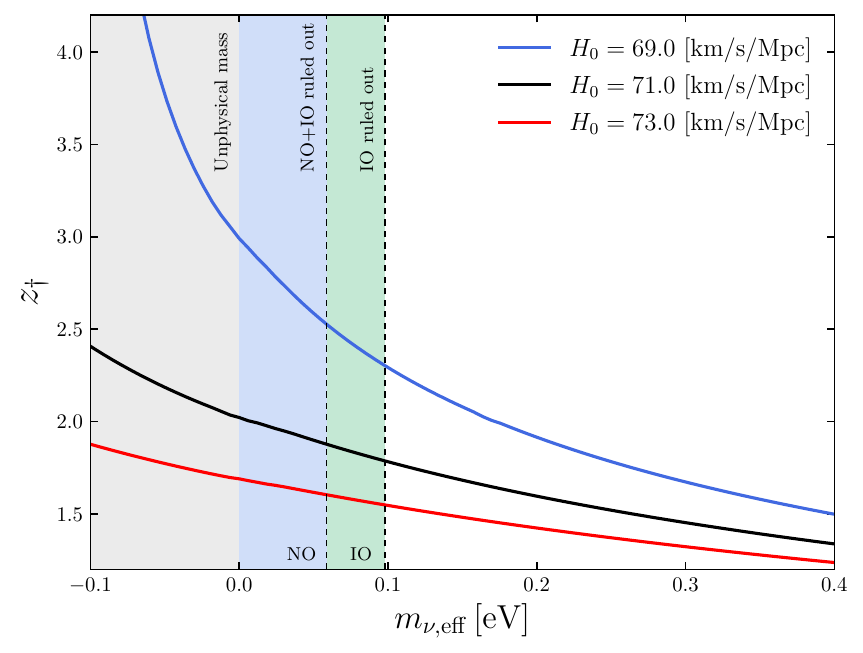}
    \caption{\label{fig:z_dag-m_nu}
    Relation between the transition redshift $z_\dagger$ and the effective neutrino-mass parameter $\sum m_{\nu,\rm eff}$ in abrupt $\Lambda_{\rm s}$CDM for selected fixed values of $H_0$. Each curve corresponds to the locus obtained by requiring consistency with the CMB acoustic scale $\theta_\star$: larger $\sum m_{\nu,\rm eff}$ requires a lower transition redshift, whereas more negative values push the model toward the $\Lambda$CDM limit at higher $z_\dagger$. The shaded vertical bands indicate the lower bounds implied by terrestrial neutrino-oscillation data for the normal and inverted orderings. The region $\sum m_{\nu,\rm eff}<0$ is shown to emphasize the phenomenological extension considered in this work.
    }
\end{figure}

\begin{figure*}[t]
    \centering
    \includegraphics[width=0.32\textwidth]{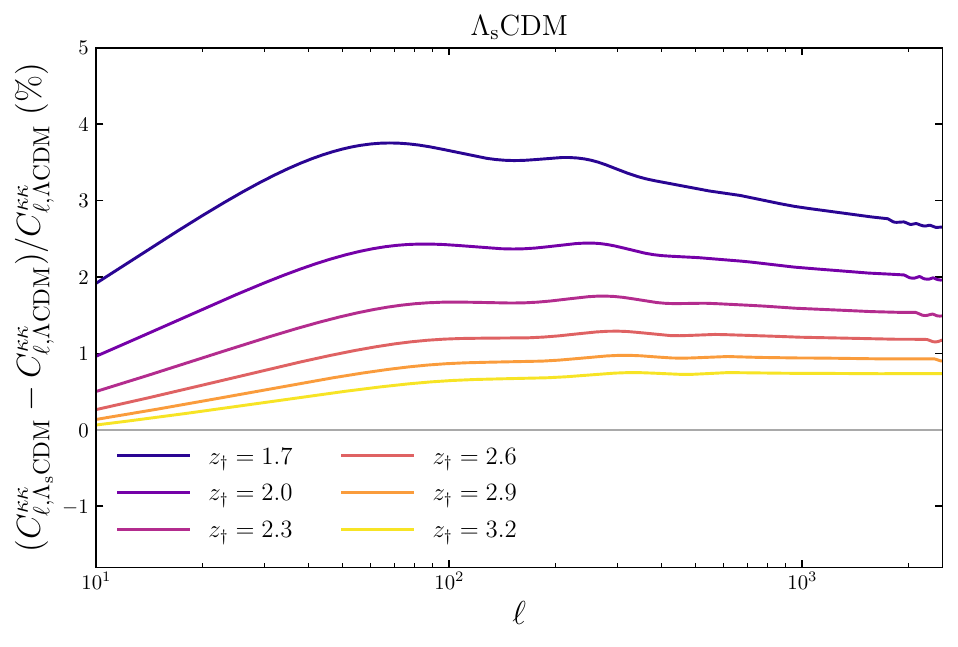}
    \includegraphics[width=0.32\textwidth]{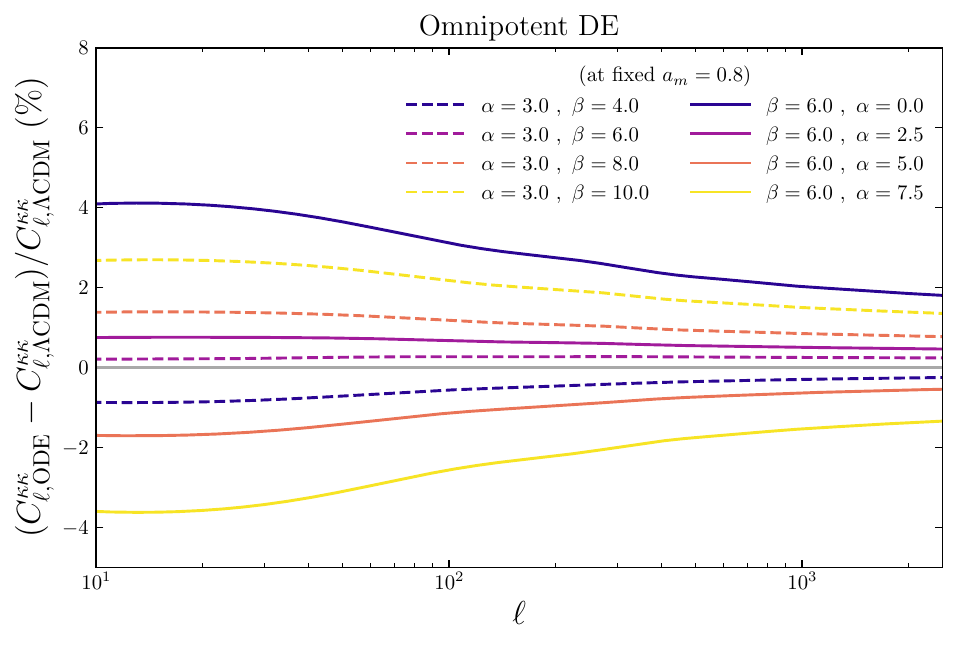}
    \includegraphics[width=0.32\textwidth]{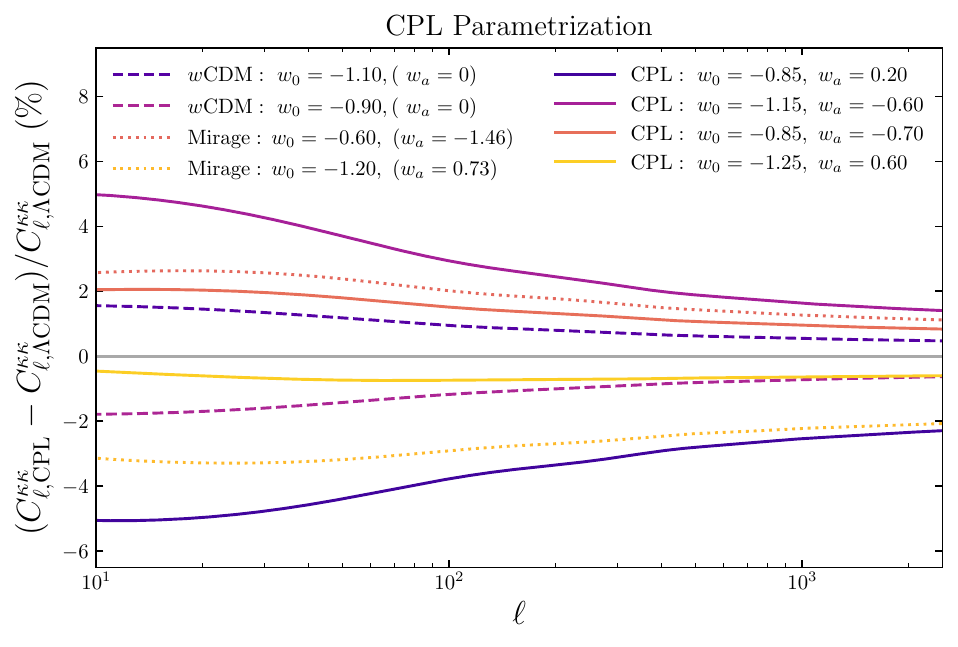}
    \caption{\label{fig:fractional_lens}
    Percentage change in the CMB lensing convergence power spectrum relative to $\Lambda$CDM,
$(C^{\kappa\kappa}_\ell-C^{\kappa\kappa}_{\ell,\Lambda{\rm CDM}})/C^{\kappa\kappa}_{\ell,\Lambda{\rm CDM}}$.
    The left panel shows $\Lambda_{\rm s}$CDM for several $z_\dagger$ values.
    The middle panel shows DMS20 ODE, varying $\beta$ at fixed $(a_m,\alpha)=(0.8,3.0)$ (dashed)
    and varying $\alpha$ at fixed $(a_m,\beta)=(0.8,6.0)$ (solid).
    The right panel shows representative $w$CDM, mirage, and CPL histories.
    Positive residuals indicate enhanced CMB lensing relative to $\Lambda$CDM. The curves are computed with \texttt{CAMB} for representative parameter choices.}
\end{figure*}

In the $\Lambda_{\rm s}\rm CDM$ model~\cite{Akarsu:2019hmw,Akarsu:2021fol,Akarsu:2022typ,Akarsu:2023mfb}, the AdS-to-dS transition is described by the sign-switching cosmological constant, $\Lambda_{\rm s}$, contributing negatively to the total energy budget of the Universe, namely $\Lambda_{\rm s}<0$ for $z>z_\dagger$, thereby causing a drop in the expansion rate. To keep $\theta_\star$, and therefore $D_{\rm M}(z_\star)$ once $r_\star$ is fixed, in~\cref{angular_diameter_dist} fixed at its precisely measured value, the post-transition expansion rate, $H(z<z_\dagger)$, must rise to compensate for the deficit in the pre-transition rate, $H(z>z_\dagger)$. Detailed observational analyses constrain the transition redshift to be around $z_\dagger \sim 2$. Therefore, at redshifts $z \lesssim 2$, $\Lambda_{\rm s}\mathrm{CDM}$ tends to yield larger $H(z)$, generally in better agreement with the indications from the low-redshift DESI BAO distance measurements. Consequently, in $\Lambda_{\rm s}$CDM extended by a free neutrino-mass parameter, the transition redshift provides an additional geometric degree of freedom: an increase in $\sum m_\nu$ can be compensated, to some extent, by shifting $z_\dagger$ to lower redshift, allowing $H_0$ to remain higher than in the corresponding $\Lambda$CDM fit. Primary CMB data alone do not sharply constrain $z_\dagger$, and allowing the neutrino mass to vary freely introduces further degeneracies which, to be broken, require complementary geometric probes such as BAO. Even so, the acoustic scale $\theta_\star$ still selects narrow slivers in the $z_\dagger$--$\sum m_{\nu,\rm eff}$ plane, as shown in~\cref{fig:z_dag-m_nu}, even when the neutrino-mass parameter $\sum m_{\nu,\rm eff}$ is extended to negative values. Some parts of these regions, where $z_\dagger \sim 1.8$, are compatible both with the local $H_0$ measurements and with the minimum mass bounds from the oscillation data. For a specific $H_0$ value, larger neutrino masses shift the transition to lower redshifts, whereas smaller or more negative $\sum m_{\nu,\rm eff}$ push $z_\dagger$ higher, making $\Lambda_{\rm s}\mathrm{CDM}$ increasingly similar to $\Lambda\mathrm{CDM}$. Therefore, a statistically significant preference for $1.7 \lesssim z_\dagger \lesssim 3$ in a joint CMB+DESI analysis would distinguish the mirror AdS-to-dS transition scenario, in which the inferred $\Omega_{\rm m}$ tends to be lower as a consequence of the elevated $H(z<z_\dagger)$~(or, equivalently, of a larger $H_0 r_{\rm d}$). This then potentially alleviates the $\Omega_{\rm m}$ mismatch, a major driver of both the shift toward negative effective neutrino masses and the exacerbation of the lensing anomaly.

From the perspective of growth, the AdS-like cosmological constant, $\Lambda_{\rm s}(z>z_\dagger)<0$, alters the growth history through the reduction of $H(z)$ at redshifts $z>z_\dagger$. The corresponding enhancement of the linear growth factor arises because the reduced expansion rate lowers the Hubble-friction term over the relevant intermediate-redshift interval, before the model returns to the positive late-time branch below $z_\dagger$. The lensing convergence power spectrum $C_\ell^{\kappa\kappa}$ is accordingly boosted, and we show the impact of varying $z_\dagger$ on $C_\ell^{\kappa\kappa}$$C_\ell^{\kappa\kappa}$ in~\cref{fig:fractional_lens}. As expected, the longer the AdS-like vacuum-energy phase persists~(or the smaller $z_\dagger$), the greater the amount of lensing excess produced relative to $\Lambda\mathrm{CDM}$. This additional lensing power can act to compensate for the suppression effect of massive neutrinos on structure growth and thereby relax the preference for lower neutrino masses.

\subsection{Neutrino Mass in Omnipotent DE}

Omnipotent dark energy~(ODE)~\cite{Adil:2023exv} is a class of DE models with non-monotonic density histories that can exhibit phantom-crossing behavior and, in parts of parameter space, access negative energy densities. Here, we focus on its specific realization, the DMS20 parametrization~\cite{DiValentino:2020naf,Adil:2023exv,Specogna:2025guo}, in which the DE density given in~\cref{eq:rhoExp} is constructed to have an extremum at $a_\mathrm{m}$, and $\alpha$, $\beta$ are the coefficients of the quadratic and cubic contributions around $a_\mathrm{m}$. For positive $\beta$, the cubic contribution $\beta(a-a_\mathrm{m})^3$ is negative for $a<a_\mathrm{m}$ and can drive $\rho_{\rm DE}$ toward a negative branch in the past while leaving it positive at late times, thereby producing, for suitable parameter choices, a sign-changing density history capable of generating a similar qualitative suppression-and-uplift pattern in $H(z)$ as in $\Lambda_\mathrm{s}$CDM~\cite{Akarsu:2019hmw,Akarsu:2021fol,Akarsu:2022typ,Akarsu:2023mfb}. The parameter space of the DMS20 parametrization is complex, yet it is still informative to vary its additional degrees of freedom in order to identify the regions relevant to our context. For fixed $a_\mathrm{m}=0.8$ and $\alpha = 3.0$, the condition for $\rho_{\rm DE}(a)$ to cross below zero at some point in the past is given by $\beta \gtrsim 5.7$. The middle panel in~\cref{fig:fractional_lens} shows that the residual lensing predicted by ODE transitions from a deficit to an excess as $\beta$ increases across this boundary. We also observe that increasing $\alpha$ while keeping the other parameters fixed amplifies the positive contribution of the quadratic term to $\rho_{\rm DE}(a)$. As a result, its effect is qualitatively opposite to that of $\beta$, as reflected in the corresponding decrease in the lensing amplitude with increasing $\alpha$ (see~\cref{fig:fractional_lens}). The parameters $\alpha$ and $\beta$ are therefore strongly degenerate. Thus, parameter combinations such as $a_\mathrm{m}=0.8$, $\alpha=3.0$, and $\beta\gtrsim5.7$ give rise to a sign-changing DE density, in a manner similar in spirit to $\Lambda_{\rm s}$CDM. Owing to its flexible background dynamics and its ability, in parts of parameter space, to enhance the lensing signal, ODE can broaden the allowed neutrino-mass range, although the data need not select the strongly negative-density branch.

\subsection{The \boldmath$w_0w_a$CDM~(or CPL) Model} \label{cpl}

The $w_0w_a\rm CDM$ model, commonly known as the Chevallier-Polarski-Linder~(CPL) parametrization~\cite{Chevallier:2000qy, Linder:2002et}, is a phenomenological extension of $\Lambda$CDM that can reproduce the behavior of several other well-known DE models. It modifies the DE equation of state, introducing a scale-factor dependence in the form
\begin{equation}
    w_{\rm DE}(a) = w_0 + w_a(1-a),
\end{equation}
where $w_0$ and $w_a$ are free parameters. With the CPL form adopted, recent DESI BAO analyses have hinted at dynamical DE with phantom-crossing behavior~\cite{DESI:2024mwx,DESI:2025zgx,DES:2025sig,Ozulker:2025ehg} (see also~\cite{DESI:2025fii,DESI:2024kob,Cortes:2024lgw,Shlivko:2024llw,Luongo:2024fww,Gialamas:2024lyw,Wang:2024dka,Ye:2024ywg,Tada:2024znt,Carloni:2024zpl,Chan-GyungPark:2024mlx,Bhattacharya:2024hep,Reboucas:2024smm,Najafi:2024qzm,Giare:2024gpk,Giare:2024ocw,Jiang:2024xnu,RoyChoudhury:2024wri,Giare:2024oil,Giare:2025pzu,Kessler:2025kju,RoyChoudhury:2025dhe,Scherer:2025esj,Wolf:2025jlc,Santos:2025wiv,Specogna:2025guo,Cheng:2025lod,Cheng:2025hug,Li:2025vuh,Lee:2025pzo,Fazzari:2025lzd,Smith:2025icl,Herold:2025hkb,Cheng:2025yue,Gokcen:2026pkq,Ishak:2025cay,Najafi:2026kxs,Yang:2026yaq}.). This preference appears to share a common origin with the tendency of the same data to push neutrino masses toward unphysically low values, associated with the so-called $\Omega_{\rm m}$ discrepancy~\cite{Elbers:2024sha,Elbers:2025vlz}. CPL-like EoS-based models can alleviate the neutrino-mass tension by reshaping the positive DE density history: phantom-like behavior at higher redshift can reduce $H(z)$ there, while the subsequent evolution can restore a larger low-redshift expansion rate~\cite{Elbers:2024sha}. As a preliminary investigation of this behavior, we vary its free parameters $w_0$ and $w_a$ as displayed in the right panel of~\cref{fig:fractional_lens}. It can be seen that, in all cases where CPL produces a larger lensing excess than $\Lambda$CDM, similarly to sign-switching dark energies, the DE equation of state exhibits phantom behavior in the past, namely $w_{\rm DE}\approx w_0 + w_a <-1$. By contrast, NPDDE defines the region in the CPL parameter space demarcated by $w_0 \geq -1$ and $w_0 + w_a \geq -1$. Its changes in both expansion and growth tend to act against the effects of positive neutrino mass, driving $\sum m_{\nu,\mathrm{eff}}$ to be even more negative. Mirage DE~\cite{Linder:2007ka}, however, occupies a more restricted trajectory in the $(w_0,w_a)$ plane. It lies along a constrained ``mirage slope'' $w_a = -3.66\,(1+w_0)$ along which smaller values of $w_0$ are accompanied by larger values of $w_a$ and $\Omega_{\rm m}$, i.e.\ a stronger mirage effect, which can partially absorb the $\Omega_{\rm m}$ discrepancy.

\begin{figure}[!ht]
    \centering
    \includegraphics[width=0.48\textwidth]{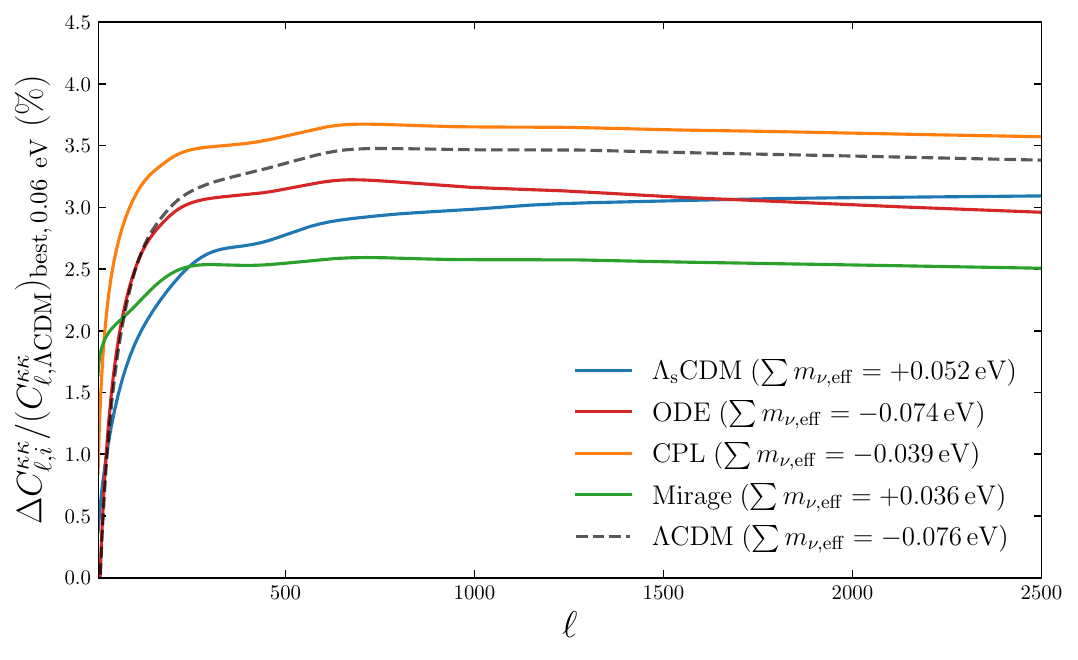}
    \includegraphics[width=0.48\textwidth]{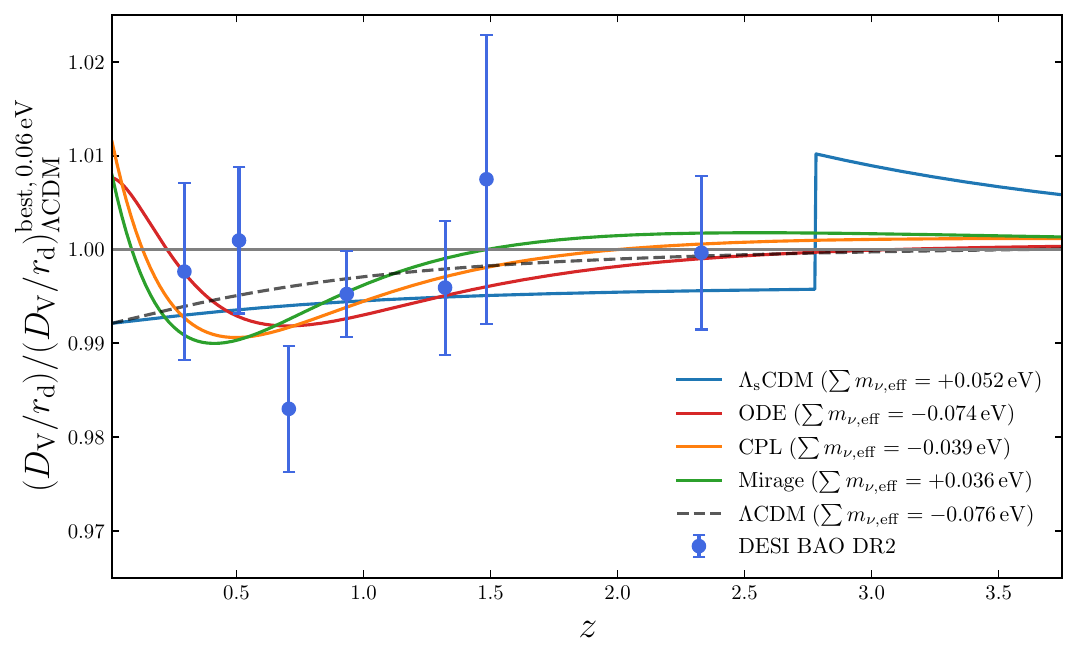}
\caption{\label{fig:lensing_bao_tradeoff}
Best-fitting results from the combined DESI DR2 BAO + CMB, where CMB denotes the primary-plus-lensing data set defined in Sec.~\ref{methods}. Top: percentage change in the CMB lensing convergence power spectrum relative to the best-fitting fixed-mass $\Lambda$CDM baseline,
$\Delta C^{\kappa\kappa}_{\ell,i}/(C^{\kappa\kappa}_{\ell,\Lambda{\rm CDM}})_{\rm best,\,0.06\,eV}$,
for $\Lambda_{\rm s}$CDM, ODE, CPL, Mirage, and free-$\sum m_{\nu,\mathrm{eff}}$ $\Lambda$CDM. Bottom: corresponding volume-averaged BAO distance ratio, $(D_{\rm V}/r_{\rm d})/(D_{\rm V}/r_{\rm d})_{\rm best,\,0.06\,eV}$, shown together with the DESI DR2 measurements. The $D_{\rm V}/r_{\rm d}$ panel is shown only as a compressed visualization; the DESI likelihood itself is built from $D_{\rm M}/r_{\rm d}$ and $D_{\rm H}/r_{\rm d}$. All best-fitting models generate a similar $\sim 3\%$--$4\%$ lensing excess, but differ in how this is reconciled with the low-redshift BAO geometry.
The more flexible EoS-based models and ODE (DMS20) density histories track the DESI distance ratios more closely, whereas $\Lambda_{\rm s}$CDM achieves a comparable lensing enhancement while shifting the inferred effective neutrino mass more decisively toward the positive region. Mirage provides an intermediate case between these two behaviors.
}
\end{figure}

Furthermore, the additional degrees of freedom in CPL --- and, more generally, in Mirage and ODE --- endow these models with greater flexibility in the late-time expansion history. This flexibility can substantially improve the fit to the low-redshift distance data; nevertheless, it comes at a cost. DESI BAO and supernova observations tightly constrain the expansion history over $0 \lesssim z \lesssim 2.3$, whereas CMB lensing is sourced more broadly over roughly $1 \lesssim z \lesssim 10$. These dark energy parametrizations therefore tend to realize their largest departures from $\Lambda$CDM precisely in the redshift range where low-redshift geometric probes are already most constraining. As a result, much of the parameter space in which an intermediate-redshift suppression of $H(z)$ could more efficiently enhance growth and lensing is disfavored by BAO and SNe, and the lensing anomaly can continue to pull $\sum m_{\nu,\rm eff}$ toward smaller, or even slightly negative, values. By contrast, the sign-switching phase of $\Lambda_{\rm s}\mathrm{CDM}$ provides a more localized and rapid deformation of the late-time expansion history, concentrating the main suppression of $H(z)$ around and above $z_\dagger$, where the BAO and supernova constraints are weaker but the lensing kernel still has substantial support. Although the model is formally defined by a sign-switched vacuum contribution for all $z>z_\dagger$, its observable effect relative to $\Lambda$CDM is concentrated in the post-recombination interval around $z\sim2$--3, while the early-time calibration remains essentially unchanged. This gives $\Lambda_{\rm s}\mathrm{CDM}$ greater leverage in shifting the posterior mean toward larger effective neutrino masses, albeit at the price of reduced flexibility in fitting the detailed low-redshift distance data.

To compare all models on equal footing and make this trade-off explicit, we adopt the effective neutrino-mass prescription presented in~\cite{Elbers:2024sha}, which defines the prediction for any observable $X$ as
\begin{equation}
\label{neutrino_mass_prescription}
\begin{aligned}
X_\theta^{\sum m_{\nu,\rm{eff}}}
&= X_\theta^{\sum m_\nu=0}+ \operatorname{sgn}\,\bigl({\scriptstyle \sum m_{\nu,\rm{eff}}}\bigr)\\ 
&\quad\quad\times\left[ X_\theta^{\left|\sum m_{\nu,\rm{eff}}\right|}
- X_\theta^{\sum m_\nu=0} \right].
\end{aligned}
\end{equation}
The above expression provides a simple continuation of the standard massive-neutrino case into the regime $\sum m_{\nu,\mathrm{eff}}<0$. Implementing this prescription in \texttt{CAMB}, we give the lensing and distance-ratio residuals in~\cref{fig:lensing_bao_tradeoff}. The top panel shows the best-fitting $\Lambda_{\rm s}\mathrm{CDM}~(z_\dagger\simeq2.78)$, ODE~($a_\mathrm{m}\simeq0.996,\,\alpha \simeq1.79,\,\beta \simeq 2.77$), CPL~($w_0 \simeq-0.82, \, w_a \simeq -0.55$), Mirage~($w_0\simeq -0.82, \, w_a=-0.66$), and free-$\sum m_{\nu,\rm eff}$ $\Lambda$CDM solutions, based on DESI DR2 BAO, CMB, and DES-Dovekie SNe data (see~\cref{methods} for details). All generate a similar lensing excess of approximately $3\%$--$4\%$, but they differ in how this excess is reconciled with the low-redshift BAO geometry. The more flexible EoS-based models and ODE density histories track the DESI distance ratios more closely, whereas $\Lambda_{\rm s}\mathrm{CDM}$ achieves a comparable lensing enhancement while shifting $\sum m_{\nu,\rm eff}$ more decisively to the positive side. Mirage represents an intermediate case, yielding a mildly positive best-fitting effective mass, but with a smaller shift than in $\Lambda_{\rm s}\mathrm{CDM}$.

\subsection{Neutrino Mass in \boldmath$\Lambda$CDM$ + \Omega_{\rm k}$} \label{omegak}

We next consider the $\Lambda$CDM+$\Omega_{\rm k}$ model. As discussed in Sec.~\ref{model}, spatial curvature provides a geometric, sign-fixed contribution to the dimensionless Friedmann equation, entering $E^2(z)$ as $\Omega_{\rm k}(1+z)^2$. For $\Omega_{\rm k}<0$, corresponding to a closed universe, this term acts as a negative contribution to $E^2(z)$ and can therefore partially mimic the phenomenological role played by a negative effective neutrino mass. If curvature is formally grouped with the positive cosmological constant, closed $\Lambda$CDM can even be viewed as a geometric analogue of a sign-changing effective contribution to the Friedmann equation~\cite{Acquaviva:2021jov}. However, this analogy remains limited: curvature is not a sign-switching DE density, carries no independent transition scale, and modifies the distance--redshift relations geometrically. Its relevance in the present context therefore comes both from its negative contribution to $E^2(z)$ for $\Omega_{\rm k}<0$ and its direct impact on the lensing geometry. Concretely, it modifies the geometry of the Universe as
\begin{equation}
D_{\rm M}(z)=f_K[\chi(z)],
\end{equation}
where
\begin{equation}
\!\!f_K(\chi)=
\begin{cases}
\dfrac{c}{H_{0}\sqrt{\Omega_{\rm k}}}\,
\sinh\!\left(\sqrt{\Omega_{\rm k}}\,\dfrac{H_0\chi}{c}\right),
& \Omega_{\rm k}>0, \\[0.9em]
\chi,
& \Omega_{\rm k}=0, \\[0.9em]
\dfrac{c}{H_{0}\sqrt{|\Omega_{\rm k}|}}\,
\sin\!\left(\sqrt{|\Omega_{\rm k}|}\,\dfrac{H_0\chi}{c}\right),
& \Omega_{\rm k}<0,
\end{cases}
\label{eq:fK}
\end{equation}
and therefore directly affects the CMB lensing potential, making it a particularly interesting case in the context of the lensing excess. The CMB lensing potential is given by
\begin{equation}
\phi(\hat{\mathbf{n}}) = -2 \int_{0}^{\chi_\star} {\rm d}\chi \,
\frac{f_K(\chi_\star-\chi)}{f_K(\chi_\star)\,f_K(\chi)} \,
\Psi [\chi \hat{\mathbf{n}}, z(\chi)],
\label{eq:lensing_potential_curved}
\end{equation}
where $\Psi [\chi \hat{\mathbf{n}}, z(\chi)]$ is the Weyl potential evaluated along the unperturbed photon geodesic at comoving radial distance $\chi$, $\chi_\star \equiv \chi(z_\star)$ is the comoving distance to last scattering, and $f_K(\chi)$ is the curvature-dependent transverse comoving-distance function defined above. Through its dependence on the comoving angular diameter distance, the lensing potential is sensitive to curvature. A closed universe tends to boost the lensing signal and therefore is typically correlated with a larger inferred lensing amplitude~\cite{DiValentino:2019qzk,DiValentino:2020hov}. Taken at face value, spatial curvature may then seem capable of absorbing part of the excess lensing signal that, in a spatially flat universe, drives the total neutrino mass toward unphysically low values.

However, this possibility is strongly limited by low-redshift geometric probes. The BAO distance-redshift relations measured by DESI, together with supernova data, constrain the late-time geometry very tightly and leave little room for departures from spatial flatness. Accordingly, once these probes are included, the data typically favor $\Omega_{\rm k}\approx0$, substantially restricting the ability of curvature to replace the role of a negative effective neutrino mass in the parameter inference. As a result, any residual curvature contribution would likely be too small to ease the tension, and the fit would still tend to push $\sum m_{\nu,\rm eff}$ toward lower values.

\section{Data and methodology} \label{methods}

In this section, we describe the data sets used in the analysis and the methodology adopted to constrain the cosmological models.
We perform the analysis using the \texttt{Cobaya} package~\cite{Torrado:2020dgo}, interfaced with a modified version of the Boltzmann solver \texttt{CLASS}~\cite{Blas:2011rf}. Dark energy perturbations are treated within the Parameterized Post-Friedmann (PPF) framework. The convergence of the MCMC chains is assessed using the Gelman-Rubin criterion~\cite{Gelman:1992zz}, requiring $R-1<0.01$, and the resulting chains are analyzed using \texttt{GetDist}~\cite{Lewis:2019xzd}. Unless otherwise stated, quoted parameter uncertainties denote marginalized $68\%$ credible intervals, while one-sided limits denote $95\%$ credible limits.

We consider the following data sets:

\begin{itemize}

\item {\bf Baryon Acoustic Oscillations (BAO):} We use BAO measurements from the second data release (DR2) of the DESI survey~\cite{DESI:2025zpo,DESI:2025zgx}. 
This dataset includes a comprehensive set of measurements of $D_{\rm M}(z)/r_{\rm d}$ and $D_{\rm H}(z)/r_{\rm d}$ over the redshift range $0.4<z<4.2$, where $D_{\rm H}(z)\equiv c/H(z)$, together with a low-redshift constraint on $D_{\rm V}(z)/r_{\rm d}$ in the range $0.1<z<0.4$, with $D_{\rm V}(z)\equiv [zD_{\rm M}^2(z)D_{\rm H}(z)]^{1/3}$. In the following, we denote this dataset as {\bf DESI}.

\item {\bf Cosmic Microwave Background (CMB):} We include low-$\ell$ temperature and polarization likelihoods, namely {\it Commander} for TT and {\it SRoll2} for EE~\cite{Pagano:2019tci}. At high multipoles, we adopt a combination of Planck and ACT data, consisting of the Planck CamSpec~\cite{Rosenberg:2022sdy} TT, TE, and EE likelihoods over the ranges $\ell \in [30,2000]$, $[30,1000]$, and $[30,600]$, respectively, together with ACT DR6 measurements over $\ell \in [2000,8500]$, $[1000,8500]$, and $[600,8500]$ for TT, TE, and EE~\cite{ACT:2025fju}. 
In addition to the primary CMB anisotropies, we include the full CMB lensing combination, which jointly uses the Planck PR4 NPIPE lensing likelihood~\cite{Carron:2022eyg}, ACT DR6 lensing reconstruction data~\cite{ACT:2023kun,ACT:2023dou}, and the SPT-3G MUSE lensing reconstruction~\cite{SPT-3G:2025zuh}. In the following, we refer to this combined dataset simply as {\bf CMB}. When needed, we distinguish between primary CMB temperature and polarization data and CMB lensing reconstruction; otherwise, the shorthand {\bf CMB} denotes the combined primary-plus-lensing data set described above.

\item {\bf Type Ia supernovae (SNIa):} We use the {\bf DES-Dovekie} supernova sample~\cite{DES:2025sig}, corresponding to the newly recalibrated DES sample and providing constraints on the late-time expansion history.

\end{itemize}

We assume flat priors on all cosmological parameters, as listed in~\cref{tab:priors}.
The baseline parameter space corresponds to the six parameters of the $\Lambda$CDM model,
$\{\Omega_{\rm b}h^2,\Omega_{\rm c}h^2,100\theta_{\rm s},\tau,n_\mathrm{s},\ln(10^{10}A_\mathrm{s})\}$,
describing respectively the physical baryon and cold dark matter densities, the angular size of the sound horizon at recombination, the optical depth to reionization, and the spectral index and amplitude of the primordial scalar power spectrum.
This parameter space is extended depending on the model under consideration. In the neutrino sector, we consider either the physical sum of neutrino masses $\sum m_\nu$ or its effective extension $\sum m_{\nu,\mathrm{eff}}$. In the latter case, the parameter is allowed to take negative values, following the perturbative prescription introduced in~\cite{Elbers:2024sha} and described further in Appendix A of~\cite{Elbers:2025vlz}, providing a continuous extension of the standard massive-neutrino framework beyond the physical boundary.
In the DE sector, we consider several parametrizations. 
For the EoS-based models, we use the CPL form $w_{\rm DE}(a)=w_0+w_a(1-a)$ as the parent parametrization. The $w$CDM model corresponds to the subcase $w_a=0$, while Mirage DE imposes the relation $w_a=-3.66(1+w_0)$. For NPDDE, we restrict the CPL parameter space so that the EoS parameter remains non-phantom over the relevant domain. For the omnipotent dark energy (ODE) framework, we vary the parameters $(\alpha,\beta,a_\mathrm{m})$ entering the parametrization of the DE density $\rho_{\rm DE}(a)$ given in~\cref{eq:rhoExp}, where $a_\mathrm{m}$ sets the location of the extremum and $(\alpha,\beta)$ control the quadratic and cubic contributions around it. For the $\Lambda_{\rm s}$CDM model, we introduce the transition redshift, $z_\dagger$, which defines the epoch at which the effective cosmological constant changes sign.
Finally, in the $\Lambda$CDM$+\Omega_{\rm k}$ model, we allow the curvature parameter $\Omega_{\rm k}$ to vary.

\begin{table}
\caption{Priors on the cosmological parameters considered in this work.}
\label{tab:priors}
\centering
    \begin{tabular}{lcc}
    \toprule
    Parameter & Default & Prior \\
    \midrule
    $\Omega_\mathrm{b}h^2$ & --- & $\mathcal{U}[0.005,0.1]$ \\
    $\Omega_\mathrm{cdm}h^2$ & --- & $\mathcal{U}[0.001,0.99]$ \\
    $100\theta_\mathrm{s}$ & --- & $\mathcal{U}[0.5,10]$ \\
    $\ln\left(10^{10}A_\mathrm{s}\right)$ & --- & $\mathcal{U}[1.61,3.91]$ \\
    $n_\mathrm{s}$ & --- & $\mathcal{U}[0.8,1.2]$ \\
    $\tau$ & --- & $\mathcal{U}[0.01,0.8]$ \\
    $\sum m_{\nu,\mathrm{eff}}$ [eV] & $0.06$ & $\mathcal{U}[-5,5]$ \\
    \midrule
    $z_\dagger$ & --- & $\mathcal{U}[1,3]$ \\
    $\alpha$ & --- & $\mathcal{U}[0,30]$ \\
    $\beta$ & --- & $\mathcal{U}[0,30]$ \\
    $a_\mathrm{m}$ & --- & $\mathcal{U}[0,1]$ \\
    $w_0$ or $w$\footnote{We require $w_0+w_a<0$ to guarantee an early period of matter domination. In the case of NPDDE, we also impose $w_0\geq -1$ and $w_0+w_a\geq -1$ so that the CPL EoS parameter remains non-phantom over $0\leq a\leq1$.} & $-1$ & $\mathcal{U}[-3,1]$ \\
    $w_a$ & $0$ & $\mathcal{U}[-3,2]$ \\
    $\Omega_{\rm k}$ & $0$ & $\mathcal{U}[-0.3,0.3]$ \\
    \bottomrule
    \end{tabular}
\end{table}

To avoid ambiguity, model names in~\cref{model} refer to the underlying late-time dark energy or geometric sector. In the parameter constraints reported below, however, labels such as $\Lambda$CDM, $\Lambda_{\rm s}$CDM, ODE, CPL, Mirage DE, NPDDE, $w$CDM, and $\Lambda$CDM+$\Omega_{\rm k}$ denote the corresponding free-$\sum m_{\nu,\mathrm{eff}}$ model cases, unless explicitly stated otherwise. The fixed-mass reference model is denoted fixed-mass $\Lambda$CDM, which assumes $\sum m_\nu=0.06\,{\rm eV}$.

We quantify the goodness of fit using the value of $\chi^2_\mathrm{MAP}\equiv -2\, \ln \mathcal{L}_\mathrm{MAP}$, where $\mathcal{L}_\mathrm{MAP}$ denotes the likelihood at the maximum a posteriori (MAP) point of the MCMC chains. The best-fitting points and corresponding $\chi^2$ values are obtained using the \texttt{BOBYQA} minimizer~\cite{Cartis:2018xum,Cartis:2018jxl}.

We compute the Bayesian evidence
\begin{equation}
    \mathcal{Z}_i \equiv \mathcal{Z}(d \, | \, \mathcal{M}_i ) = \int {\rm d}\theta \, p(d \, | \, \theta, \mathcal{M}_i)\, p(\theta \, | \, \mathcal{M}_i),
\end{equation}
using the \texttt{harmonic} code~\cite{mcewen2023machinelearningassistedbayesian},\footnote{\href{https://github.com/astro-informatics/harmonic}{github.com/astro-informatics/harmonic}} and cross-checked the results with the \texttt{MCEvidence} package~\cite{Heavens:2017hkr,Heavens:2017afc},\footnote{\href{https://github.com/yabebalFantaye/MCEvidence}{github.com/yabebalFantaye/MCEvidence}} using a \texttt{Cobaya}-compatible wrapper.\footnote{\href{https://github.com/williamgiare/wgcosmo/tree/main}{github.com/williamgiare/wgcosmo/tree/main}} Here, $\mathcal{L}=p(d \, | \, \theta,\mathcal{M}_i)$ denotes the likelihood and $p(\theta \, | \, \mathcal{M}_i)$ the prior distribution, while $d$, $\theta$, and $\mathcal{M}_i$ represent the data, model parameters, and the model, respectively. We find consistent results between the two implementations and report the results obtained with \texttt{harmonic}.

Model comparison is performed through the Bayes factor
\begin{equation}
    \ln \mathcal{B}_{ij} = \ln \mathcal{Z}_i - \ln \mathcal{Z}_j,
\end{equation}
defined with respect to a reference model $j$. Unless otherwise stated, the reference model is the standard fixed-mass $\Lambda$CDM model, for which $\sum m_\nu=0.06\,{\rm eV}$. Positive values of $\ln\mathcal{B}_{ij}$ favor model $i$, while negative values indicate a preference for this reference model. We interpret $\mathcal{B}_{ij}$ using the Jeffreys scale~\cite{Trotta:2008qt}: $|\ln\mathcal{B}_{ij}|<1$ corresponds to inconclusive evidence, $1 \lesssim |\ln\mathcal{B}_{ij}| \lesssim 2.5$ to weak evidence, $2.5 \lesssim |\ln\mathcal{B}_{ij}| \lesssim 5$ to moderate evidence, and $|\ln\mathcal{B}_{ij}| \gtrsim 5$ to strong evidence.

\section{Results} \label{results}

\begin{figure}[!t]
    \includegraphics[width=0.48\textwidth]{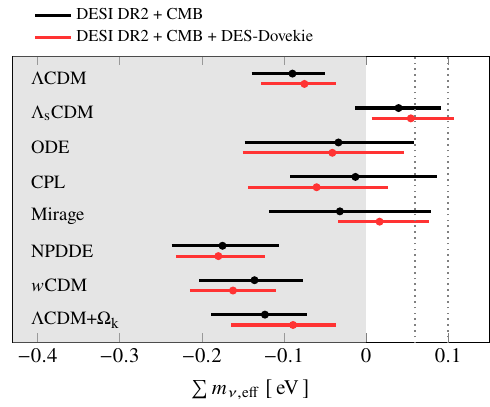}\vspace{-1em}
\caption{\label{fig:whiskers1}
Summary of the constraints on the effective neutrino mass parameter, $\sum m_{\nu,\mathrm{eff}}$, from DESI DR2 BAO + CMB alone (black) and combined with DES-Dovekie SNe (red), for the set of models considered in this work. Points and horizontal bars indicate marginalized posterior means and $68\%$ credible intervals. The vertical dotted lines denote the lower bounds from neutrino oscillation experiments for the normal and inverted orderings. The figure highlights that $\Lambda_{\rm s}$CDM is the only model in the current set that yields a $68\%$ credible interval for $\sum m_{\nu,\mathrm{eff}}$ entirely within the positive range for the full data combination.
}
\end{figure}

\begin{figure*}[ht!]
    \subfloat{
        \includegraphics[scale=0.85]{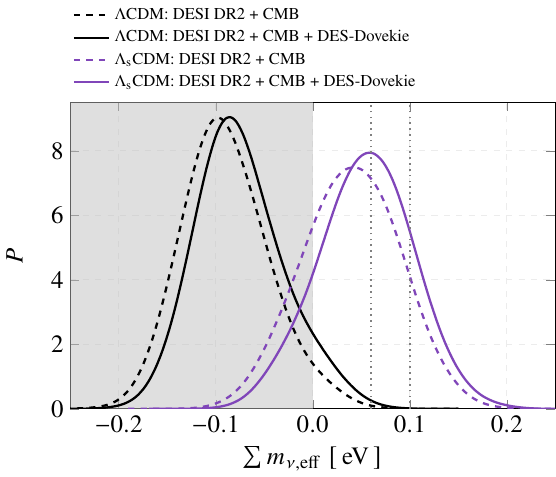}
    }
    \subfloat{
        \includegraphics[scale=0.85]{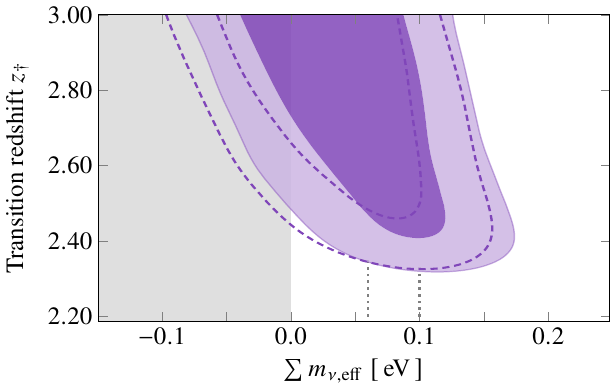}
    }\vspace{-1em}
\caption{\label{fig:Mnu_posteriors}
Posterior distributions of the effective neutrino-mass parameter, $\sum m_{\nu,\mathrm{eff}}$, for free-$\sum m_{\nu,\mathrm{eff}}$ $\Lambda$CDM and $\Lambda_{\rm s}$CDM, obtained from DESI DR2 BAO + CMB alone (dashed) and combined with DES-Dovekie SNe (solid). The right panel shows the $68\%$ and $95\%$ posterior credible regions in the $\sum m_{\nu,\mathrm{eff}}$--$z_\dagger$ plane for $\Lambda_{\rm s}$CDM, where $z_\dagger$ denotes the mirror AdS-to-dS transition redshift. The joint posteriors are truncated by the upper prior bound on $z_\dagger$. The vertical dotted lines in both panels indicate the lower bounds on the physical neutrino-mass sum from oscillation experiments for the normal and inverted mass orderings.
}
\end{figure*}

\begin{figure*}[ht!]
    \subfloat{
        \includegraphics[scale=0.55]{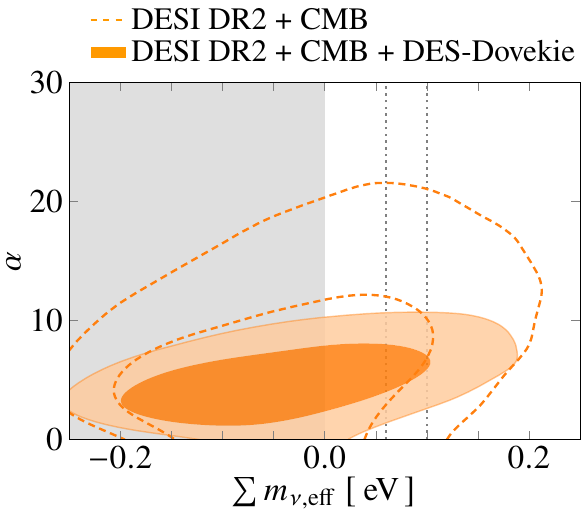}
    }
    \subfloat{
        \includegraphics[scale=0.55]{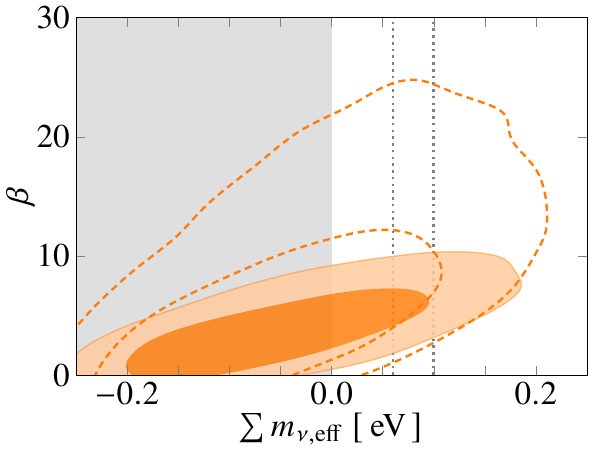}
    }
    \subfloat{
        \includegraphics[scale=0.55]{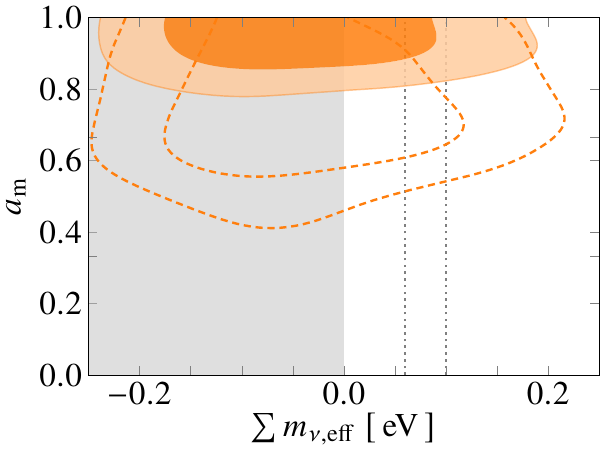}
    }\vspace{-1em}
    \caption{\label{fig:Mnu_posteriors_ode}
Contour plots showing the $68\%$ and $95\%$ posterior credible regions for the effective neutrino-mass parameter, $\sum m_{\nu,\mathrm{eff}}$, and the three parameters of the omnipotent dark energy model, $(\alpha,\beta,a_{\mathrm m})$, for DESI DR2 BAO + CMB alone (dashed) and combined with DES-Dovekie SNe (filled). The figure shows that positive neutrino masses are allowed over large regions of the ODE parameter space, enabled by the positive correlation between $\sum m_{\nu,\mathrm{eff}}$, $\alpha$, and $\beta$. The vertical dotted lines indicate the lower bounds on the neutrino mass from oscillation experiments for the normal and inverted orderings.
}
\end{figure*}

\begin{figure}[!t]
    \centering
    \includegraphics[width=0.48\textwidth]{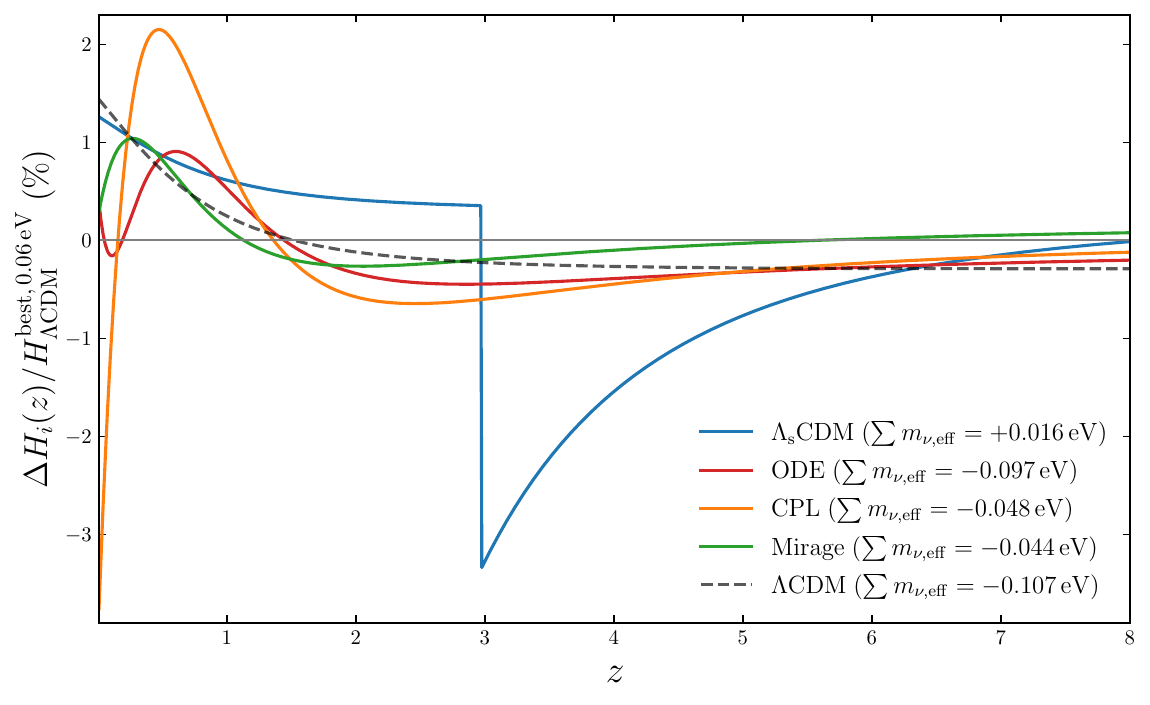}
    \includegraphics[width=0.48\textwidth]{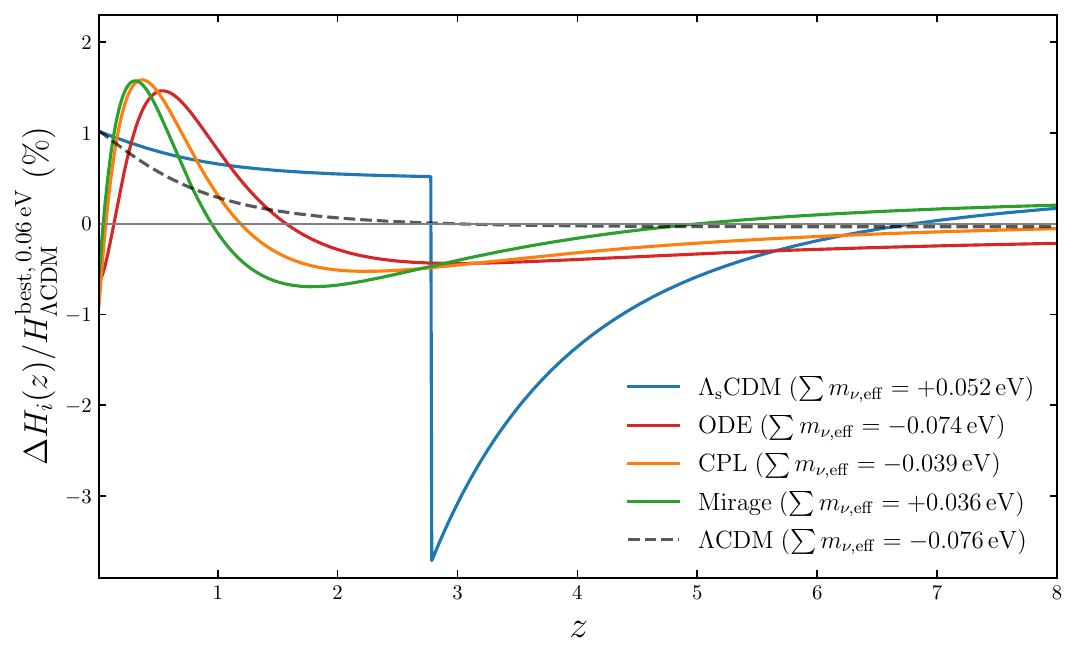}
\caption{\label{fig:DeltaHoverH} Fractional deviation in the expansion rate,
$\Delta H_i(z)/H_{\Lambda\mathrm{CDM}}^{\mathrm{best},\,0.06\,\mathrm{eV}}$,
where $\Delta H_i(z)\equiv H_i(z)-H_{\Lambda\mathrm{CDM}}^{\mathrm{best},\,0.06\,\mathrm{eV}}$, for the best-fitting free-$\sum m_{\nu,\mathrm{eff}}$ $\Lambda_{\rm s}$CDM, ODE, CPL, Mirage DE, and $\Lambda$CDM model cases relative to the best-fitting fixed-mass $\Lambda$CDM reference with $\sum m_\nu=0.06\,{\rm eV}$. The sharp feature in the $\Lambda_{\rm s}$CDM curve reflects the abrupt sign switch at the corresponding best-fitting transition redshift. The top panel shows the best-fitting curves obtained from the combined dataset consisting of DESI DR2 BAO, CMB temperature and polarization data from \textit{Planck} and ACT, and CMB lensing reconstruction from \textit{Planck}, ACT, and SPT. The bottom panel shows the same quantity for the extended data combination in which the DES-Dovekie Type Ia supernova compilation is added to DESI DR2 BAO + CMB.
}
\end{figure}

\begin{figure}[!t]
    \centering
    \includegraphics[width=0.48\textwidth]{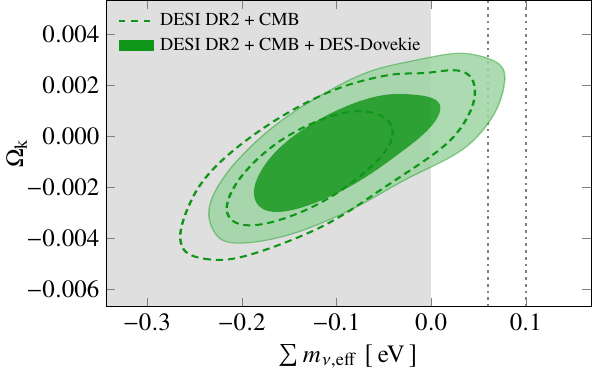}
\caption{\label{fig:Mnu_Omega_k}
Posterior credible regions at $68\%$ and $95\%$ for the effective neutrino mass, $\sum m_{\nu,\mathrm{eff}}$, and curvature, $\Omega_{\rm k}$, for DESI DR2 BAO + CMB alone (dashed) and combined with DES-Dovekie SNe (filled). Due to the correlation between the two parameters, positive neutrino masses can be recovered for $\Omega_{\rm k}>0$. This positive-mass region is distinct from the closed-curvature branch that provides the formal negative contribution to $E^2(z)$ discussed in the text. The vertical dotted lines indicate the lower bounds on the neutrino mass from oscillation experiments for the normal and inverted orderings.
}
\end{figure}

We now summarize the parameter constraints presented in~\cref{tab:neutrino_constraints,tab:neutrino_constraints_des}, and in~\cref{fig:whiskers1,fig:Mnu_posteriors,fig:Mnu_posteriors_ode,fig:DeltaHoverH,fig:Mnu_Omega_k}. Unless explicitly labeled as fixed-mass $\Lambda$CDM, all model names in this section refer to the corresponding free-$\sum m_{\nu,\mathrm{eff}}$ model cases. We use three complementary diagnostics to compare the models: whether the inferred effective neutrino mass is shifted away from the unphysical negative region, whether the fit to the data improves, and whether the improvement is achieved through the kind of late-time deformation identified in~\cref{anatomy}. This distinction is important because the model that gives the largest decrease in the best-fitting $\chi^2$ need not be the model that most transparently replaces the phenomenological role played by $\sum m_{\nu,\mathrm{eff}}<0$ in $\Lambda$CDM.

\begin{table*}

\caption{
Cosmological parameter constraints for nine model cases: fixed-mass $\Lambda$CDM, with $\sum m_\nu=0.06\,{\rm eV}$, and eight free-$\sum m_{\nu,\mathrm{eff}}$ cases: $\Lambda$CDM, $\Lambda_{\rm s}$CDM, ODE, CPL, Mirage DE, NPDDE, $w$CDM, and $\Lambda$CDM+$\Omega_{\rm k}$ from DESI DR2 BAO combined with CMB temperature and polarization data from Planck and ACT, and CMB lensing reconstruction from Planck, ACT, and SPT. We report marginalized posterior means with $68\%$ credible intervals, except for $z_\dagger$ in $\Lambda_{\rm s}$CDM and the $\beta$ parameter in ODE, for which one-sided $95\%$ credible bounds are quoted.
}
\centering
\resizebox{\linewidth}{!}{
    \begin{tabular}{lccccccccc}
    \toprule
    \textbf{Parameter} & $\mathbf{\Lambda}$\textbf{CDM}$\,+\,$fixed$\,\sum m_\nu$ & $\mathbf{\Lambda}$\textbf{CDM} & $\mathbf{\Lambda_\mathrm{s}}$\textbf{CDM} & \textbf{ODE} & \textbf{CPL} & \textbf{Mirage} & \textbf{NPDDE} & $\mathbf{w}$\textbf{CDM} & $\mathbf{\Lambda}$\textbf{CDM}$+\mathbf{\Omega_{\rm k}}$ \\
    \midrule
    $100\Omega_\mathrm{b}h^2$ & $2.2493\pm 0.0096$ & $2.2459\pm 0.0095$ & $2.2419\pm 0.0095$ & $2.2457\pm 0.0099$ & $2.245\pm 0.010$ & $2.245\pm 0.010$ & $2.2488\pm 0.0097$ & $2.2477\pm 0.0096$ & $2.249\pm 0.010$ \\
    $\Omega_\mathrm{cdm}h^2$ & $0.11781\pm 0.00059$ & $0.11880\pm 0.00065$ & $0.11960\pm 0.00066$ & $0.11893\pm 0.00079$ & $0.11913\pm 0.00078$ & $0.11910\pm 0.00078$ & $0.11832\pm 0.00071$ & $0.11851\pm 0.00072$ & $0.1182\pm 0.0011$ \\
    $100\theta_\mathrm{s}$ & $1.04186\pm 0.00020$ & $1.04177\pm 0.00020$ & $1.04169\pm 0.00020$ & $1.04175\pm 0.00020$ & $1.04173\pm 0.00021$ & $1.04174\pm 0.00020$ & $1.04178\pm 0.00020$ & $1.04177\pm 0.00020$ & $1.04181\pm 0.00021$ \\
    $\ln\left(10^{10}A_\mathrm{s}\right)$ & $3.072^{+0.011}_{-0.012}$ & $3.051^{+0.011}_{-0.013}$ & $3.054^{+0.010}_{-0.012}$ & $3.051^{+0.011}_{-0.013}$ & $3.052^{+0.010}_{-0.013}$ & $3.052^{+0.011}_{-0.013}$ & $3.048^{+0.010}_{-0.012}$ & $3.049^{+0.011}_{-0.013}$ & $3.047^{+0.011}_{-0.013}$ \\
    $n_\mathrm{s}$ & $0.9735\pm 0.0028$ & $0.9715\pm 0.0028$ & $0.9696\pm 0.0029$ & $0.9712\pm 0.0030$ & $0.9708\pm 0.0030$ & $0.9707\pm 0.0029$ & $0.9725\pm 0.0030$ & $0.9721\pm 0.0029$ & $0.9728\pm 0.0033$ \\
    $\tau$ & $0.0710^{+0.0060}_{-0.0068}$ & $0.0605^{+0.0056}_{-0.0067}$ & $0.0608^{+0.0054}_{-0.0065}$ & $0.0604^{+0.0055}_{-0.0065}$ & $0.0604^{+0.0053}_{-0.0067}$ & $0.0606^{+0.0054}_{-0.0065}$ & $0.0598^{+0.0055}_{-0.0065}$ & $0.0600^{+0.0054}_{-0.0066}$ & $0.0596^{+0.0055}_{-0.0066}$ \\
    $\sum m_{\nu,\mathrm{eff}}/$eV & $0.06$ (fixed) & $-0.090^{+0.040}_{-0.050}$ & $0.040\pm 0.051$ & $-0.034^{+0.092}_{-0.11}$ & $-0.013^{+0.10}_{-0.079}$ & $-0.032^{+0.11}_{-0.086}$ & $-0.175^{+0.069}_{-0.062}$ & $-0.136^{+0.059}_{-0.067}$ & $-0.123^{+0.052}_{-0.065}$ \\
    $z_\dagger$ & --- & --- & $ > 2.41$ & --- & --- & --- & --- & --- & ---\\
    $\alpha$ & --- & --- & --- & $3.35^{+0.83}_{-3.2}$ & --- & --- & --- & --- & ---\\
    $\beta$ & --- & --- & --- & $<19.1$ & --- & --- & --- & --- & ---\\
    $a_\mathrm{m}$ & --- & --- & --- & $0.849\pm 0.082$ & --- & --- & --- & --- & ---\\
    $w_0$ or $w$ & --- & --- & --- & --- & $-0.59^{+0.23}_{-0.26}$ & $-0.89^{+0.13}_{-0.17}$ & $-1.01^{+0.11}_{-0.080}$ & $-0.957\pm 0.045$ & ---\\
    $w_a$ & --- & --- & --- & --- & $-1.15^{+0.83}_{-0.68}$ & $-0.42^{+0.63}_{-0.46}$ & $0.18^{+0.12}_{-0.26}$ & --- & ---\\
    $\Omega_{\rm k}$ & --- & --- & --- & --- & --- & --- & --- & --- & $-0.0011\pm 0.0015$\\
    \midrule
    $h$ & $0.6823\pm 0.0025$ & $0.6911\pm 0.0038$ & $0.6903\pm 0.0037$ & $0.678^{+0.01}_{-0.0068}$ & $0.650\pm 0.022$ & $0.683^{+0.013}_{-0.011}$ & $0.687^{+0.012}_{-0.015}$ & $0.6825\pm 0.0096$ & $0.6911\pm 0.0038$ \\
    $\Omega_\mathrm{m}$ & $0.3028\pm 0.0034$ & $0.2938\pm 0.0043$ & $0.2989\pm 0.0043$ & $0.283^{+0.026}_{-0.015}$ & $0.336^{+0.024}_{-0.027}$ & $0.303^{+0.012}_{-0.015}$ & $0.295^{+0.012}_{-0.011}$ & $0.2997\pm 0.0074$ & $0.2918^{+0.0047}_{-0.0054}$ \\
    $\sigma_8$ & $0.8170\pm 0.0046$ & $0.8437\pm 0.0096$ & $0.8247\pm 0.0095$ & $0.828^{+0.015}_{-0.013}$ & $0.803^{+0.023}_{-0.026}$ & $0.833\pm 0.018$ & $0.847^{+0.013}_{-0.016}$ & $0.840\pm 0.010$ & $0.848\pm 0.011$ \\
    $S_8$ & $0.8208\pm 0.0064$ & $0.8349\pm 0.0078$ & $0.8232\pm 0.0078$ & $0.803^{+0.042}_{-0.012}$ & $0.848\pm 0.011$ & $0.8369\pm 0.0081$ & $0.8386\pm 0.0093$ & $0.8390\pm 0.0090$ & $0.8363\pm 0.0079$ \\
    $r_\mathrm{d}/$Mpc & $147.55\pm 0.17$ & $147.33\pm 0.18$ & $147.16\pm 0.18$ & $147.30\pm 0.21$ & $147.25\pm 0.20$ & $147.26\pm 0.21$ & $147.44\pm 0.20$ & $147.39\pm 0.20$ & $147.47\pm 0.26$ \\
    \midrule
    $\Delta\chi^2_\mathrm{MAP}$ & $0.0$ & $-10.68$ & $-9.099$ & $-11.81$ & $-13.19$ & $-11.57$ & $-12.13$ & $-11.99$ & $-11.86$ \\
    $\ln B$ & $0.0$ & $0.66$ & $-1.83$ & $-3.91$ & $-2.35$ & $-1.53$ & $-4.08$ & $-2.41$ & $-4.11$ \\
    \bottomrule
    \end{tabular}
}
\label{tab:neutrino_constraints}
\end{table*}

\begin{table*}
\caption{
Cosmological parameter constraints for nine model cases: fixed-mass $\Lambda$CDM, with $\sum m_\nu=0.06\,{\rm eV}$, and eight free-$\sum m_{\nu,\mathrm{eff}}$ cases: $\Lambda$CDM, $\Lambda_{\rm s}$CDM, ODE, CPL, Mirage DE, NPDDE, $w$CDM, and $\Lambda$CDM+$\Omega_{\rm k}$ from DESI DR2 BAO combined with Type Ia supernovae from the DES-Dovekie compilation, CMB temperature and polarization data from Planck and ACT, and CMB lensing reconstruction from Planck, ACT, and SPT. We report marginalized posterior means with $68\%$ credible intervals, except for $z_\dagger$ in $\Lambda_{\rm s}$CDM and the $a_{\rm m}$ parameter in ODE, for which one-sided $95\%$ credible bounds are quoted.}
\centering
\resizebox{\linewidth}{!}{
    \begin{tabular}{lccccccccc}
    \toprule
    \textbf{Parameter} & $\mathbf{\Lambda}$\textbf{CDM}$\,+\,$fixed$\,\sum m_\nu$ & $\mathbf{\Lambda}$\textbf{CDM} & $\mathbf{\Lambda_\mathrm{s}}$\textbf{CDM} & \textbf{ODE} & \textbf{CPL} & \textbf{Mirage} & \textbf{NPDDE} & $\mathbf{w}$\textbf{CDM} & $\mathbf{\Lambda}$\textbf{CDM}$+\mathbf{\Omega_{\rm k}}$ \\
    \midrule
    $100\Omega_\mathrm{b}h^2$ & $2.2482\pm 0.0094$ & $2.2452\pm 0.0096$ & $2.2414\pm 0.0095$ & $2.2455\pm 0.0098$ & $2.2458\pm 0.0098$ & $2.2429\pm 0.0096$ & $2.2493\pm 0.0096$ & $2.2489\pm 0.0096$ & $2.246\pm 0.010$ \\
    $\Omega_\mathrm{cdm}h^2$ & $0.11796\pm 0.00058$ & $0.11889\pm 0.00065$ & $0.11971\pm 0.00066$ & $0.11894\pm 0.00077$ & $0.11888\pm 0.00076$ & $0.11931\pm 0.00066$ & $0.11825\pm 0.00069$ & $0.11833\pm 0.00068$ & $0.1187\pm 0.0010$ \\
    $100\theta_\mathrm{s}$ & $1.04185\pm 0.00020$ & $1.04176\pm 0.00020$ & $1.04168\pm 0.00020$ & $1.04175\pm 0.00020$ & $1.04175\pm 0.00020$ & $1.04172\pm 0.00020$ & $1.04179\pm 0.00021$ & $1.04179\pm 0.00020$ & $1.04178\pm 0.00021$ \\
    $\ln\left(10^{10}A_\mathrm{s}\right)$ & $3.071\pm 0.012$ & $3.051^{+0.011}_{-0.013}$ & $3.054^{+0.011}_{-0.013}$ & $3.051^{+0.011}_{-0.013}$ & $3.051^{+0.011}_{-0.012}$ & $3.053^{+0.010}_{-0.013}$ & $3.047^{+0.011}_{-0.012}$ & $3.049^{+0.010}_{-0.013}$ & $3.050^{+0.011}_{-0.014}$ \\
    $n_\mathrm{s}$ & $0.9731\pm 0.0028$ & $0.9712\pm 0.0029$ & $0.9695\pm 0.0029$ & $0.9711\pm 0.0030$ & $0.9713\pm 0.0030$ & $0.9703\pm 0.0028$ & $0.9725\pm 0.0029$ & $0.9725\pm 0.0029$ & $0.9718\pm 0.0034$ \\
    $\tau$ & $0.0703^{+0.0060}_{-0.0066}$ & $0.0606^{+0.0056}_{-0.0069}$ & $0.0608^{+0.0054}_{-0.0067}$ & $0.0604^{+0.0055}_{-0.0066}$ & $0.0603^{+0.0056}_{-0.0066}$ & $0.0607^{+0.0054}_{-0.0066}$ & $0.0596^{+0.0054}_{-0.0064}$ & $0.0600^{+0.0055}_{-0.0066}$ & $0.0602^{+0.0056}_{-0.0068}$ \\
    $\sum m_{\nu,\mathrm{eff}}/$eV & $0.06$ (fixed) & $-0.075^{+0.039}_{-0.053}$ & $0.055\pm 0.050$ & $-0.041^{+0.087}_{-0.11}$ & $-0.060\pm 0.083$ & $0.017^{+0.060}_{-0.051}$ & $-0.180^{+0.057}_{-0.052}$ & $-0.162\pm 0.054$ & $-0.089^{+0.052}_{-0.075}$\\
    $z_\dagger$ & --- & --- & $ > 2.40$ & --- & --- & --- & --- & --- & ---\\
    $\alpha$ & --- & --- & --- & $2.33^{+0.87}_{-1.3}$ & --- & --- & --- & --- & ---\\
    $\beta$ & --- & --- & --- & $4.1^{+1.7}_{-3.1}$ & --- & --- & --- & --- & ---\\
    $a_\mathrm{m}$ & --- & --- & --- & $>0.895$ & --- & --- & --- & --- & ---\\
    $w_0$ or $w$ & --- & --- & --- & --- & $-0.830\pm 0.072$ & $-0.822\pm 0.074$ & $-0.930^{+0.044}_{-0.040}$ & $-0.932\pm 0.028$ & ---\\
    $w_a$ & --- & --- & --- & --- & $-0.46\pm 0.30$ & $-0.65\pm 0.27$ & $0.026^{+0.057}_{-0.13}$ & --- & ---\\
    $\Omega_{\rm k}$ & --- & --- & --- & --- & --- & --- & --- & --- & $-0.0005\pm 0.0015$\\
    \midrule
    $h$ & $0.6816\pm 0.0024$ & $0.6894\pm 0.0037$ & $0.6888\pm 0.0036$ & $0.6784\pm 0.0059$ & $0.6731\pm 0.0067$ & $0.6776\pm 0.0059$ & $0.6761\pm 0.0064$ & $0.6773\pm 0.0061$ & $0.6892\pm 0.0036$ \\
    $\Omega_\mathrm{m}$ & $0.3037\pm 0.0033$ & $0.2958\pm 0.0042$ & $0.3009\pm 0.0042$ & $0.3046\pm 0.0066$ & $0.3107\pm 0.0072$ & $0.3092\pm 0.0069$ & $0.3037\pm 0.0059$ & $0.3033\pm 0.0052$ & $0.2951^{+0.0047}_{-0.0052}$ \\
    $\sigma_8$ & $0.8170\pm 0.0045$ & $0.8409\pm 0.0097$ & $0.8222\pm 0.0094$ & $0.829\pm 0.012$ & $0.826\pm 0.011$ & $0.825\pm 0.011$ & $0.837\pm 0.010$ & $0.8370\pm 0.0094$ & $0.843\pm 0.011$ \\
    $S_8$ & $0.8220\pm 0.0064$ & $0.8349\pm 0.0079$ & $0.8233\pm 0.0079$ & $0.8353\pm 0.0087$ & $0.8405\pm 0.0082$ & $0.8373\pm 0.0078$ & $0.8424\pm 0.0083$ & $0.8414\pm 0.0083$ & $0.8359\pm 0.0080$ \\
    $r_\mathrm{d}/$Mpc & $147.53\pm 0.17$ & $147.31\pm 0.18$ & $147.14\pm 0.19$ & $147.30\pm 0.20$ & $147.31\pm 0.20$ & $147.23\pm 0.18$ & $147.45\pm 0.19$ & $147.43\pm 0.19$ & $147.36\pm 0.26$ \\
    \midrule
    $\Delta\chi^2_\mathrm{MAP}$ & $0.0$ & $-8.104$ & $-7.041$ & $-15.00$ & $-16.49$ & $-14.85$ & $-15.13$ & $-14.62$ & $-9.043$ \\
    $\ln B$ & $0.0$ & $-0.21$ & $-2.14$ & $-4.79$ & $-2.06$ & $-0.175$ & $-3.44$ & $-1.28$ & $-5.18$ \\
    \bottomrule
    \end{tabular}
}
\label{tab:neutrino_constraints_des}
\end{table*}

In this language, previous analyses suggest that shifting the inferred neutrino mass toward positive values generally requires a reduction of the effective DE contribution over part of the post-recombination expansion history, roughly at $z\gtrsim1$~\cite{Yang:2026yaq}. The question addressed here is therefore not merely whether additional late-time freedom broadens the neutrino-mass posterior, but whether it supplies the specific negative effective contribution to the expansion history that the free-$\sum m_{\nu,\mathrm{eff}}$ $\Lambda$CDM fit is using as a proxy.

The first result is that the neutrino mass anomaly is clearly present in the baseline DESI DR2 BAO + CMB combination, where CMB denotes the primary-plus-lensing data set defined in~\cref{methods}. In the free-$\sum m_{\nu,\mathrm{eff}}$ $\Lambda$CDM case, the posterior peaks in the unphysical region and gives
\begin{equation}
\sum m_{\nu,\mathrm{eff}} =
-0.090^{+0.040}_{-0.050}\,{\rm eV},
\end{equation}
with $H_0=69.11\pm0.38\,{\rm km\,s^{-1}\,Mpc^{-1}}$ and $S_8=0.8349\pm0.0078$. Relative to the fixed-mass $\Lambda$CDM reference, the decrease in the best-fitting $\chi^2$ is substantial, while the Bayesian evidence remains nearly unchanged. Thus, once the physical boundary $\sum m_\nu\ge0$ is removed, the data do not simply broaden toward low masses: the preferred region moves into the negative domain. In this sense, free-$\sum m_{\nu,\mathrm{eff}}$ $\Lambda$CDM is best understood as a diagnostic model, exposing the magnitude and direction of the mismatch between current data and the standard cosmological model.

The comparison across the extended late-time models shows that not all additional freedom is equally effective. In the baseline combination, the ODE, CPL, and Mirage DE cases soften the anomaly relative to free-$\sum m_{\nu,\mathrm{eff}}$ $\Lambda$CDM, but only partially: their marginalized means remain close to, or slightly below, zero. By contrast, NPDDE, $w$CDM, and $\Lambda$CDM+$\Omega_{\rm k}$ leave the preferred value clearly in the negative region. This already shows that the data are not merely selecting generic late-time flexibility. They are selecting a more specific deformation, one that can mimic the effect of a negative contribution to the expansion history over the redshift range relevant to the DESI--CMB--lensing mismatch.

Against this background, free-$\sum m_{\nu,\mathrm{eff}}$ $\Lambda_{\rm s}$CDM behaves differently. For the same baseline data combination, we find
\begin{equation}
\sum m_{\nu,\mathrm{eff}} =
0.040\pm0.051\,{\rm eV},
\quad
z_\dagger > 2.41 \,\, (95\%),
\end{equation}
with $H_0=69.03\pm0.37\,{\rm km\,s^{-1}\,Mpc^{-1}}$ and $S_8=0.8232\pm0.0078$. Thus, relative to free-$\sum m_{\nu,\mathrm{eff}}$ $\Lambda$CDM, the sign-switching model preserves the elevated value of $H_0$ selected by the DESI-driven late-time geometry, moves the effective neutrino mass back toward the physical region, and avoids the larger $S_8$ associated with the free negative-mass fit. This is precisely the pattern expected if the fit is replacing one effective negative contribution, previously carried by $\sum m_{\nu,\mathrm{eff}}<0$, with another supplied by the negative branch of the sign-switching vacuum energy. In this restricted mechanism-level sense, $\Lambda_{\rm s}$CDM provides the cleanest explicit realization of the late-time deformation identified in~\cref{anatomy}.

This should not be confused with model selection in the usual statistical sense. In the baseline table, some of the more flexible alternatives improve the best-fitting likelihood as much as, or more than, $\Lambda_{\rm s}$CDM, while the Bayesian evidence penalizes the enlarged parameter volume of the extended models. The significance of free-$\sum m_{\nu,\mathrm{eff}}$ $\Lambda_{\rm s}$CDM here is therefore not that it is the global statistical winner, but that the $\Lambda_{\rm s}$CDM model most directly translates the negative-effective-mass diagnostic into a physical late-time contribution with the required sign structure.

The inclusion of DES-Dovekie SNe sharpens this interpretation rather than overturning it. Once the supernova compilation is added to DESI DR2 BAO, CMB temperature and polarization data, and CMB lensing reconstruction, the free-$\sum m_{\nu,\mathrm{eff}}$ $\Lambda$CDM anomaly weakens but does not disappear:
\begin{equation}
\sum m_{\nu,\mathrm{eff}}
=
-0.075^{+0.039}_{-0.053}\,{\rm eV}.
\end{equation}
Thus, additional low-redshift distance information reduces slightly the pull toward negative effective mass, but does not restore a physical neutrino-mass inference within $\Lambda$CDM. By contrast, $\Lambda_{\rm s}$CDM shifts further toward the physical side,
\begin{equation}
\sum m_{\nu,\mathrm{eff}}
=
0.055\pm0.050\,{\rm eV},
\quad
z_\dagger>2.40 \,\, (95\%),
\end{equation}
with $H_0=68.88\pm0.36\,{\rm km\,s^{-1}\,Mpc^{-1}}$ and $S_8=0.8233\pm0.0079$. In this case, the $68\%$ credible interval lies entirely within the positive range of $\sum m_{\nu,\mathrm{eff}}$ and close to the lower limit implied by neutrino oscillation data. The preferred transition scale also remains stable relative to the baseline analysis, indicating that the shift in neutrino mass is tied to the same sign-switching deformation rather than to an accidental shift induced by the supernova data.

This behavior is summarized in~\cref{fig:whiskers1}. The free-$\sum m_{\nu,\mathrm{eff}}$ $\Lambda$CDM result remains the clearest phenomenological diagnostic of the anomaly, while $\Lambda_{\rm s}$CDM is the only model in the present set for which the supernova-augmented $68\%$ credible interval is entirely in the positive region. The ODE, CPL, and Mirage DE cases soften the anomaly but do not remove it as decisively at the level of the marginalized posterior. NPDDE, $w$CDM, and $\Lambda$CDM+$\Omega_{\rm k}$ remain less successful in this respect, leaving the preferred mass in the negative region or close to it. This hierarchy reinforces the conclusion that the relevant issue is not simply the number of late-time degrees of freedom, but whether the model realizes the specific sign and redshift structure needed to replace the role of $\sum m_{\nu,\mathrm{eff}}<0$.

The supernova-augmented table also makes clear that restoring a positive effective neutrino mass is not identical to maximizing the raw goodness of fit. With DES-Dovekie included, several EoS-based models with sign-preserving $\rho_{\rm DE}$ achieve larger decreases in the best-fitting $\chi^2$ than $\Lambda_{\rm s}$CDM, and the Bayesian evidence penalizes the additional parameter volume of most extended models. Among the EoS-based cases with sign-preserving $\rho_{\rm DE}$, Mirage DE is the closest competitor, yielding
\begin{equation}
\sum m_{\nu,\mathrm{eff}}
=
0.017^{+0.060}_{-0.051}\,{\rm eV}.
\end{equation}
This shift is weaker than in $\Lambda_{\rm s}$CDM, but it shows that a restricted EoS trajectory can partially mimic the required late-time deformation once supernova distances are included.

The direct comparison between free-$\sum m_{\nu,\mathrm{eff}}$ $\Lambda$CDM and $\Lambda_{\rm s}$CDM is shown in~\cref{fig:Mnu_posteriors}. The left panel illustrates that the sign-switching model shifts the effective-mass posterior away from the negative region for both data combinations, with the shift becoming especially clear after the inclusion of DES-Dovekie SNe. The right panel shows the corresponding joint posterior in the $\sum m_{\nu,\mathrm{eff}}$--$z_\dagger$ plane. Within the adopted prior range, the data prefer a transition at the upper end of the allowed interval, yielding a lower bound $z_\dagger\gtrsim2.4$ rather than a sharply localized measurement of the transition redshift. The posterior truncation by the prior should therefore be interpreted with care: the result indicates a preference for a sufficiently early sign switch within the explored rapid-transition regime, not a precise determination of $z_\dagger$.
In this work, we adopt a commonly used uniform prior
$z_\dagger\in[1,3]$; see, e.g., Ref.~\cite{Yadav:2024duq} and footnote~3 therein. This choice ensures that the transition occurs within the intermediate-redshift window, where the DE contribution remains non-negligible and can therefore have observable leverage on the DESI BAO distances and CMB-lensing response, as discussed in~\cref{anatomy}. A much broader prior extending to larger $z_\dagger$ would allow the fit to explore a near-$\Lambda$CDM regime, with the formal $\Lambda$CDM limit recovered as $z_\dagger\to\infty$, and would therefore be expected to shift the $\sum m_{\nu,\mathrm{eff}}$ posterior back toward smaller values.

The ODE constraints provide a useful stress test of this interpretation. As shown in~\cref{fig:Mnu_posteriors_ode}, positive values of $\sum m_{\nu,\mathrm{eff}}$ are allowed over extended regions of the DMS20 parameter space, in particular along directions correlated with the density-shape parameters $\alpha$ and $\beta$. This is consistent with the density-level nature of the model: unlike EoS-based parametrizations with sign-preserving $\rho_{\rm DE}$, ODE has enough freedom to access a negative-density branch for suitable parameter choices. However, the posterior weight does not concentrate strongly in the regions where this negative branch acts as a clean replacement for $\sum m_{\nu,\mathrm{eff}}<0$. Consequently, the marginalized neutrino-mass posterior is shifted only weakly toward the physical side. In this respect, ODE has the structural capacity to realize the desired mechanism, but current data do not select that capacity decisively.

The same distinction is visible when comparing ODE with the EoS-based benchmarks whose $\rho_{\rm DE}$ is sign-preserving. CPL and Mirage DE can partially soften the anomaly by reshaping the positive DE density history and redistributing the late-time distance budget. As noted above, Mirage DE is the closest EoS-based competitor in this sense, especially once supernova distances are included. However, because their $\rho_{\rm DE}$ does not cross zero in the EoS-based construction considered here, they explicitly do not implement a negative-density branch. NPDDE and $w$CDM are more restrictive still: by limiting the allowed evolution of the EoS, they leave less freedom to reproduce the required suppression-and-uplift pattern in $H(z)$ and therefore tend to leave $\sum m_{\nu,\mathrm{eff}}$ on the negative side. The curvature case provides a complementary geometric control. Although $\Omega_{\rm k}<0$ can enter $E^2(z)$ as a negative, sign-fixed contribution, the combined BAO, CMB, and supernova geometry leaves little room for curvature to replace the role played by a negative effective neutrino mass.

\Cref{fig:DeltaHoverH} provides a useful consistency check of the mechanism discussed in~\cref{anatomy}. The figure shows the fractional deviation of the best-fitting expansion histories from the fixed-mass $\Lambda$CDM reference, for the baseline data combination and after adding DES-Dovekie SNe. The qualitative pattern is clear. The models that most directly address the anomaly suppress $H(z)$ over the intermediate-redshift range relevant for CMB lensing and restore a larger expansion rate at lower redshift. The different models realize this pattern in different ways. In $\Lambda_{\rm s}$CDM, the effect is tied to a single transition scale and is therefore sharply organized in redshift. In the EoS-based and ODE cases, the deformation is less localized and more distributed across the low- and intermediate-redshift expansion history. This explains why several models can generate comparable lensing enhancement or fit the low-redshift distances well, while still differing substantially in how efficiently they move the inferred neutrino mass into the physical region.

Taken together, the results point to a clear empirical hierarchy. Free-$\sum m_{\nu,\mathrm{eff}}$ $\Lambda$CDM provides the sharpest phenomenological diagnostic of the anomaly: once the physical boundary is removed, the data prefer the negative side. EoS-based models with sign-preserving $\rho_{\rm DE}$ can improve the fit or soften the anomaly by broadening degeneracies, while ODE has the structural freedom to access a negative-density branch, but its posterior support is not concentrated in the regions that would act as a clean replacement for $\sum m_{\nu,\mathrm{eff}}<0$. Spatial curvature provides a useful geometric control. Although closed $\Lambda$CDM can be formally grouped with $\Lambda$ to produce a sign-changing effective contribution to $E^2(z)$, this is a geometric analogue rather than a DE density transition, and the combined BAO, CMB, and supernova geometry leaves too little room for it to replace the role played by $\sum m_{\nu,\mathrm{eff}}<0$. This is also reflected directly in the inferred neutrino mass: the $\Lambda$CDM+$\Omega_{\rm k}$ model gives
\begin{equation}
\sum m_{\nu,\mathrm{eff}}
=
-0.123^{+0.052}_{-0.065}\,{\rm eV}
\end{equation}
for DESI DR2 BAO+CMB, and
\begin{equation}
\sum m_{\nu,\mathrm{eff}}
=
-0.089^{+0.052}_{-0.075}\,{\rm eV}
\end{equation}
after adding DES-Dovekie SNe. Thus, in the present fits, allowing curvature does not move the effective neutrino mass back toward the physical region. This is visible directly in~\cref{fig:Mnu_Omega_k}: the closed branch, $\Omega_{\rm k}<0$, correlates with more negative values of $\sum m_{\nu,\mathrm{eff}}$, whereas positive effective masses are recovered only on the open branch, $\Omega_{\rm k}>0$. This conclusion can be further quantified using the formal crossing scale introduced in~\cref{model}. A finite crossing exists only on the closed branch, $\Omega_{\rm k}<0$, whose posterior weight is $P(\Omega_{\rm k}<0)=0.79$ for DESI DR2 BAO+CMB and $P(\Omega_{\rm k}<0)=0.64$ after adding DES-Dovekie SNe. Conditioning on this branch, and quoting $68\%$ credible intervals, the derived posterior for DESI DR2 BAO+CMB gives
\begin{equation}
z_{\rm cross}\big|_{\Omega_{\rm k}<0}
=
28.2^{+0.6}_{-17},
\end{equation}
while after adding DES-Dovekie SNe we find
\begin{equation}
z_{\rm cross}\big|_{\Omega_{\rm k}<0}
=
32.6^{+1.5}_{-20}.
\end{equation}
The finite curvature crossing is therefore neither present throughout the posterior nor sharply localized near the $z\sim2$ regime relevant to the sign-switching models; it disappears on the open branch, $\Omega_{\rm k}>0$, and is pushed to arbitrarily high redshift as $\Omega_{\rm k}\to0^{-}$. The curvature analogue remains a useful geometric control, but it does not realize the specific sign- and redshift-structured vacuum-energy transition supplied by $\Lambda_{\rm s}$CDM.

By contrast, $\Lambda_{\rm s}$CDM is the most direct density-level realization of the mechanism identified in~\cref{anatomy}: it replaces the negative effective neutrino-mass contribution with an explicit negative vacuum-energy branch, while preserving the elevated $H_0$ preferred by the DESI-driven geometry and avoiding the enhanced $S_8$ characteristic of the free negative-mass $\Lambda$CDM fit. This does not make $\Lambda_{\rm s}$CDM the statistically preferred model within the present free-$\sum m_{\nu,\mathrm{eff}}$ comparison, but it does make it the cleanest mechanism-level translation of the negative-effective-mass diagnostic among the models considered here.

\section{Discussion} \label{discussion}

The results of this work suggest that the neutrino mass tension should be interpreted primarily as a diagnostic of missing late-time cosmological structure. Once the physical prior $\sum m_\nu\geq0$ is relaxed, the DESI DR2 BAO + CMB fit moves toward $\sum m_{\nu,\mathrm{eff}}<0$ because the minimal $\Lambda$CDM framework has no explicit component capable of supplying the negative effective contribution to the expansion history that the data appear to prefer.

Before interpreting the model comparison, it is important to specify the scope of the statistical statements. The goodness-of-fit and evidence comparisons reported in Sec.~\ref{results} refer to the model cases analyzed in this work: the fixed-mass $\Lambda$CDM reference, with $\sum m_\nu=0.06\,{\rm eV}$, and the corresponding late-time extensions in which the effective neutrino-mass parameter $\sum m_{\nu,\mathrm{eff}}$ is allowed to vary. These comparisons should therefore not be read as statements about the separate fixed-neutrino-mass comparison between $\Lambda_{\rm s}$CDM and $\Lambda$CDM, or between ODE and $\Lambda$CDM. Fixed-neutrino-mass analyses of $\Lambda_{\rm s}$CDM and related sign-switching scenarios have found fits comparable to or better than $\Lambda$CDM, and, in several data combinations, can alleviate cosmological tensions~\cite{Akarsu:2021fol,Akarsu:2022typ,Akarsu:2023mfb,Akarsu:2024qsi,Akarsu:2024eoo}. Similarly, the DMS20 model of ODE has been shown to provide a useful phenomenological framework for accessing negative-density DE histories and alleviating tensions in specific data combinations~\cite{Adil:2023exv,Specogna:2025guo}. The question addressed here is more specific: \textit{once the negative-effective-mass diagnostic is made available, which late-time sector most cleanly removes the need for $\sum m_{\nu,\mathrm{eff}}<0$?}

This interpretation is also consistent with earlier studies of neutrino-sector extensions of $\Lambda_{\rm s}$CDM, in which physical parameters such as $N_{\rm eff}$ and $\sum m_\nu$ were varied rather than analytically continued to negative effective values. Those analyses found that the late-time mirror AdS-to-dS transition can remain compatible with standard neutrino properties, including values of $N_{\rm eff}$ close to the Standard Model prediction and neutrino-mass bounds compatible with oscillation constraints, while still alleviating cosmological tensions for appropriate data combinations~\cite{Yadav:2024duq}. The present work asks a complementary question: \textit{when the effective neutrino-mass parameter is allowed to reveal the sign and magnitude of the mismatch, can the required negative contribution be supplied by the late-time cosmological sector instead?}

Within this free-$\sum m_{\nu,\mathrm{eff}}$ comparison, there are two distinct notions of success. The first is statistical: whether a model decreases the best-fitting $\chi^2$ and how strongly it is penalized by its additional parameter volume in the Bayesian evidence. The second is physical: whether the model replaces the phenomenological role of $\sum m_{\nu,\mathrm{eff}}<0$ with a concrete and interpretable late-time mechanism. These criteria need not identify the same model. In particular, once supernova data are included, several EoS-based parametrizations whose $\rho_{\rm DE}$ is sign-preserving achieve larger decreases in the best-fitting $\chi^2$ than $\Lambda_{\rm s}$CDM. Nevertheless, among the models considered here, $\Lambda_{\rm s}$CDM provides the most direct density-level realization of the required mechanism: it shifts the inferred effective neutrino mass toward the physical region through an explicit negative branch in the vacuum-energy sector.

This distinction is essential for interpreting the status of $\Lambda_{\rm s}$CDM in the present analysis. We do not find that $\Lambda_{\rm s}$CDM is the global statistical winner within the full set of free-$\sum m_{\nu,\mathrm{eff}}$ extensions, nor do the present data uniquely require a sign-switching cosmological constant. Rather, the result is more precise. We find that the type of late-time physics capable of solving the neutrino mass tension is already nontrivially constrained. A successful explanation that operates at the level of the expansion history must suppress $H(z)$ over the intermediate-redshift range relevant for CMB lensing, restore a larger expansion rate at lower redshift so as to remain compatible with DESI BAO and CMB geometry, and generate the required lensing response with the appropriate redshift and scale dependence. Within the present model set, $\Lambda_{\rm s}$CDM realizes this pattern in the most transparent minimal way.

The behavior of the comparison models clarifies what is, and is not, sufficient. EoS-based parametrizations with sign-preserving $\rho_{\rm DE}$ such as CPL and Mirage DE can redistribute the late-time distance budget and alleviate the neutrino-mass tension. Mirage DE is the closest EoS-based competitor once supernova distances are included. The main distinction with $\Lambda_\mathrm{s}$CDM is that their $\rho_{\rm DE}$ remains positive and does not cross zero. ODE, by contrast, has the structural freedom to access a negative-density branch, but in the present free-$\sum m_{\nu,\mathrm{eff}}$ analysis, the posterior does not concentrate in the regions where that branch acts as a clean replacement for $\sum m_{\nu,\mathrm{eff}}<0$. This does not contradict earlier fixed-mass ODE analyses, where the negative-density feature can play an important role in alleviating cosmological tensions; rather, it shows that the DESI--CMB--lensing neutrino-mass diagnostic considered here selects a more specific region of parameter space than is favored by the marginalized ODE posterior.

Spatial curvature provides a complementary geometric control. For $\Omega_{\rm k}<0$, the curvature term itself enters $E^2(z)$ as a negative sign-fixed contribution; if formally grouped with the positive cosmological constant, the combination can even behave as a sign-changing effective source. This makes closed $\Lambda$CDM a useful geometric analogue, but not a genuine sign-switching DE density. In the present fits, however, a finite crossing exists only on the closed branch and is not robustly selected by the posterior: substantial posterior weight remains on models with no finite crossing, and when the crossing exists, it is broadly distributed at high redshift rather than localized near the $z\sim2$ regime relevant to the sign-switching models. The combined BAO, CMB, and supernova geometry therefore leaves too little room for this geometric analogue to replace the role played by negative effective neutrino mass. This illustrates an important point: neither generic late-time flexibility, nor the mere availability of a negative contribution to the Friedmann equation, is sufficient by itself. The relevant contribution must have the right sign, redshift dependence, and compatibility with the geometric and lensing information.

The inclusion of DES-Dovekie SNe further sharpens this picture. Supernova distances weaken the neutrino mass tension in free-$\sum m_{\nu,\mathrm{eff}}$ $\Lambda$CDM but do not remove it. They also improve the performance of EoS-based models with sign-preserving $\rho_{\rm DE}$ in the global distance fit, which explains why raw goodness of fit and physical replacement separate more clearly in the supernova-augmented analysis. This reinforces one of the main messages of the paper: the model that best fits the low-redshift distance data need not be the model that most directly captures the physical role played by negative effective neutrino mass in $\Lambda$CDM.

The posterior for the transition redshift in $\Lambda_{\rm s}$CDM should also be interpreted with appropriate caution. Within the adopted prior range, the data prefer a transition toward the upper end of the allowed interval, giving a lower bound around $z_\dagger\gtrsim2.4$ rather than a sharply localized measurement of the transition redshift. Thus, the present analysis identifies a preference for a sufficiently early sign switch within the rapid-transition regime explored here, but a wider-prior study would be needed to determine how far this preference extends and whether the posterior eventually turns over. This prior sensitivity does not undermine the qualitative mechanism, but it does limit how strongly one should interpret the numerical value of $z_\dagger$.

The lensing response provides a particularly important discriminator. The models compared here can produce similar integrated enhancements of the CMB lensing amplitude, but they need not do so in the same way. What matters is not only the total amount of additional lensing, but where in redshift and multipole space the enhancement is generated. Future analyses that exploit the scale dependence and redshift sensitivity of the lensing signal more directly may therefore distinguish between sign-switching density histories, EoS-based deformations whose $\rho_{\rm DE}$ is sign-preserving, non-monotonic density models, curvature, and possible alternatives outside the model set considered here.

From a broader perspective, the main outcome of this work is not only a set of parameter constraints, but a reframing of the neutrino-mass tension. The preference for $\sum m_{\nu,\mathrm{eff}}<0$ should not be regarded simply as a pathology to be removed by imposing a prior. It can instead be used as a diagnostic of the direction in which the data pull the minimal cosmological model. In the present case, that direction points toward a sign- and redshift-structured modification of the late-time expansion history, rather than toward arbitrary additional freedom. Whether the missing contribution is ultimately best described by a sign-switching cosmological constant, a more general non-monotonic density history, or some other late-time mechanism remains open. 
In this limited but important sense, $\Lambda_{\rm s}$CDM serves as a useful benchmark for translating the neutrino mass tension into a concrete late-time cosmological mechanism, even if it is not uniquely selected by the current data.

\section{Conclusions} \label{sec:conc}

In this work, we have performed a systematic investigation of alternative dark energy (DE) models in the context of the neutrino mass tension. We use the effective neutrino mass parameter, $\sum m_{\nu,\mathrm{eff}}$, as a diagnostic in order to reveal the magnitude and direction of the mismatch between the data and the standard cosmological framework. Within free-$\sum m_{\nu,\mathrm{eff}}$ $\Lambda$CDM, the DESI DR2 BAO + CMB data yield values in the unphysical negative effective-mass range,
$\sum m_{\nu,\mathrm{eff}}=-0.090^{+0.040}_{-0.050}\,{\rm eV}$ for the baseline data combination and
$\sum m_{\nu,\mathrm{eff}}=-0.075^{+0.039}_{-0.053}\,{\rm eV}$ after adding DES-Dovekie SNe. The improvement in fit obtained in this way should not be interpreted as a physical success of $\Lambda$CDM. On the contrary, it shows that the standard late-time sector can improve its agreement with the data only by exploiting an unphysical direction in parameter space. In this sense, $\sum m_{\nu,\mathrm{eff}}<0$ acts as a phenomenological proxy for a missing negative effective contribution to the late-time expansion history.

The comparison with extended late-time sectors shows that this role is not reproduced by arbitrary additional freedom. EoS-based parametrizations of DE whose $\rho_{\rm DE}$ is sign-preserving, such as CPL, Mirage DE, NPDDE, and $w$CDM, can improve the fit by modifying the expansion history. In the case of CPL and Mirage, they can also shift the $\sum m_{\nu,\mathrm{eff}}$ posterior toward positive masses, recovering the minimum value under the normal mass ordering, $\sum m_\nu=\SI{0.059}{\eV}$, to within $1\sigma$. However, their DE densities do not cross zero and therefore they do not realize a negative-density branch. The DMS20 model of ODE has the structural freedom to access such a branch, but in the present data combinations the posterior does not concentrate in the region where this behavior acts as a clean replacement for $\sum m_{\nu,\mathrm{eff}}<0$. Spatial curvature provides a complementary geometric control: for $\Omega_{\rm k}<0$ it can enter $E^2(z)$ as a negative sign-fixed contribution, and when formally grouped with $\Lambda$ it gives a simple geometric analogue of a sign-changing effective source. However, the corresponding crossing exists only on the closed branch and is not robustly localized in the intermediate-redshift regime selected by the neutrino-mass diagnostic; it is absent on the open branch and is pushed to high redshift as $\Omega_{\rm k}\to0^{-}$. Thus, curvature does not realize the required sign- and redshift-structured contribution to the expansion history.

By contrast, $\Lambda_{\rm s}$CDM provides the most direct density-level realization of the mechanism identified in this paper. Its negative vacuum-energy branch supplies an explicit negative contribution to the expansion history above the transition, with observational leverage concentrated around the intermediate-redshift regime after recombination. In the baseline data combination, $\Lambda_{\rm s}$CDM shifts the inferred effective neutrino mass to
$\sum m_{\nu,\mathrm{eff}}=0.040\pm0.051\,{\rm eV}$, and after adding DES-Dovekie SNe it shifts it to
$\sum m_{\nu,\mathrm{eff}}=0.055\pm0.050\,{\rm eV}$, with $z_\dagger>2.4$ at $95\%$ credibility. The posterior support for $z_\dagger$ lies toward the upper end of the adopted prior range, so this result should be interpreted as evidence for a sufficiently early transition within the rapid-transition regime explored here, rather than as a precise measurement of the transition redshift.

It is important to emphasize the scope of this conclusion. The statistical comparisons in this paper are made between the fixed-mass $\Lambda$CDM reference and the corresponding free-$\sum m_{\nu,\mathrm{eff}}$ late-time extensions. Therefore, the fact that $\Lambda_{\rm s}$CDM is not the global statistical winner within this diagnostic comparison should not be read as a statement about separate fixed-neutrino-mass comparisons of $\Lambda_{\rm s}$CDM or ODE with standard $\Lambda$CDM. In the present analysis, goodness of fit and Bayesian evidence serve as important consistency checks, but they are not the primary diagnostic of success. The central question is whether the underlying late-time sector can remove the need for the fit to enter the unphysical negative-mass domain while remaining compatible with the data.

From this perspective, the significance of $\Lambda_{\rm s}$CDM is that it reveals the type of mechanism that can alleviate the neutrino mass tension: when the effective neutrino-mass direction is opened, it is, among the models considered here, the one that most cleanly reduces the need for $\sum m_{\nu,\mathrm{eff}}<0$ by replacing that phenomenological contribution with a concrete negative branch in the vacuum-energy sector. The broader lesson is therefore not simply that more late-time freedom is needed, but that the data appear to point toward a particular kind of freedom: a sign- and redshift-structured deformation of the expansion history, suppressing $H(z)$ over the redshift range relevant for CMB lensing and growth, while restoring a larger expansion rate at lower redshifts that is compatible with DESI BAO distances and CMB geometry.

More broadly, our results point to a three-way connection between the cosmological neutrino-mass tension, the discrepancy between DESI BAO measurements and the $\Lambda$CDM predictions inferred from CMB data, and the structure of the CMB lensing signal. The negative-effective-mass direction exposes the mismatch; the DESI BAO distances constrain the required low-redshift geometry; and the CMB lensing information selects how the intermediate-redshift expansion history and growth of structure must respond. This connection is the reason why the problem cannot be resolved by arbitrary parameter freedom alone. What is required is a late-time sector with the right sign, redshift dependence, and lensing response.

Whether the missing contribution is ultimately best described by a sign-switching cosmological constant, a more general non-monotonic density history, or another late-time mechanism remains open. Future work should therefore focus not only on global parameter constraints, but also on observables that can resolve the redshift and scale dependence of the effect. In particular, forthcoming full-shape galaxy-clustering analyses, improved CMB lensing measurements, increasingly precise low-redshift distance data, as well as future BAO distance measurements at higher redshifts, should be able to test whether the required lensing enhancement and expansion-history deformation arise in the manner predicted by $\Lambda_{\rm s}$CDM-like sign-switching scenarios, by more distributed non-monotonic density histories, or by some different late-time mechanism.

In this limited but important sense, $\Lambda_{\rm s}$CDM serves as a useful benchmark for DESI-era analyses of neutrino mass and dark energy physics. Whether or not it is the final answer, it isolates a concrete mechanism by which the neutrino mass tension can be translated into a physically interpretable late-time cosmological deformation. The tension should therefore not be viewed merely as a pathology to be removed by imposing a prior, but as a useful diagnostic of the magnitude and structure of the mismatch between current data and the minimal cosmological model.

\begin{acknowledgments}
W.E. acknowledges STFC Consolidated Grant ST/X001075/1. \"{O}.A.\ acknowledges the support of the Turkish Academy of Sciences in the scheme of the Outstanding Young Scientist Award (T\"{U}BA-GEB\.{I}P). E.D.V. is supported by a Royal Society Dorothy Hodgkin Research Fellowship. This article is based upon work from the COST Action CA21136 ``Addressing observational tensions in cosmology with systematics and fundamental physics'' (CosmoVerse), supported by COST (European Cooperation in Science and Technology). This work used the DiRAC@Durham facility managed by the Institute for Computational Cosmology on behalf of the STFC DiRAC HPC Facility (www.dirac.ac.uk). The equipment was funded by BEIS capital funding via STFC capital grants ST/K00042X/1, ST/P002293/1, ST/R002371/1 and ST/S002502/1, Durham University and STFC operations grant ST/R000832/1. DiRAC is part of the National e-Infrastructure.
\end{acknowledgments}

\bibliographystyle{apsrev4-2_mod_yearfix} 
\bibliography{apssamp}

\end{document}